\documentclass[%
 aip,
 amsmath,amssymb,
 reprint,%
]{revtex4-1}

\usepackage{graphicx}
\usepackage{dcolumn}
\usepackage{bm}
\usepackage{multirow}

\usepackage[utf8]{inputenc}
\usepackage[T1]{fontenc}
\usepackage{mathptmx}
\usepackage{etoolbox}
\usepackage{xcolor}

\makeatletter
\def\@email#1#2{%
 \endgroup
 \patchcmd{\titleblock@produce}
  {\frontmatter@RRAPformat}
  {\frontmatter@RRAPformat{\produce@RRAP{*#1\href{mailto:#2}{#2}}}\frontmatter@RRAPformat}
  {}{}
}%
\makeatother
\begin{document}

\preprint{AIP/123-QED}

\title[Power- and time-dependent equivalent circuit models for waveform-selective metasurfaces]{Power- and time-dependent equivalent circuit models for waveform-selective metasurfaces with varying electromagnetic responses to repeated pulses at the same frequency}
\author{Ryuho Miyamoto}
\author{Hiroki Wakatsuchi}%
 \email{wakatsuchi.hiroki@nitech.ac.jp}
\affiliation{ 
Department of Engineering, Graduate School of Engineering, Nagoya Institute of Technology, Gokiso-cho, Showa, Nagoya, Aichi, 466-8555, Japan
}%

\date{\today}

\begin{abstract}
Waveform-selective metasurfaces offer unprecedented control over electromagnetic waves on the basis of pulse width. However, existing circuit models fail to capture the power-dependent behaviors of these metasurfaces, thereby limiting their use in practical applications. Here, for the first time, we present analytical equivalent circuit models that accurately predict both power- and time-dependent responses by incorporating voltage-dependent diode resistance through the Maclaurin series and Wright omega functions. As a result, the variations in the input power and time domain are effectively predicted theoretically. Moreover, our concept is successfully extended to different types of waveform-selective metasurfaces and increasingly complex scenarios, including repeated pulses and nonresonant frequencies. Thus, our equivalent circuit approach can readily explain and quantify the electromagnetic behaviors of waveform-selective metasurfaces. This strategy provides a high degree of control for addressing complex electromagnetic problems by leveraging pulse width as a tuning parameter, even at a fixed frequency.
\end{abstract}

\maketitle

\section{Introduction}
Electromagnetic fields and related phenomena can readily be tailored by engineered structures called metamaterials and metasurfaces.\cite{smithDNG1D, smithDNG2D2, EBGdevelopment, calozBook, MTMbookEngheta} Their subwavelength unit cells can artificially be designed to respond to incident electromagnetic waves, generating exotic properties such as negative or zero refractive indices,\cite{smithDNG1D, smithDNG2D2, ziolkowski2004propagation, liberal2017near} high surface impedances,\cite{EBGdevelopment} and abrupt phase changes.\cite{yu2011light, yu2014flat} These unique properties can be used to address a range of issues and realize novel application devices, including perfect lenses that overcome diffraction limits,\cite{pendryperfetLenses, fangSuperlens, grbic2008near} cloaking,\cite{pendryCloaking, schurig2006metamaterial, enghetaCloaking} suppression of electromagnetic interference,\cite{mtmAbsPRLpadilla, My1stAbsPaper} and analog computing for solving complicated equations at the speed of light.\cite{silva2014performing, mohammadi2019inverse, zangeneh2021analogue}

In particular, metasurfaces are intensively studied in the domain of wireless communications because of the growing demand for next-generation wireless communications. Over the past two decades, metamaterials and metasurfaces have been integrated into the design of antennas to markedly reduce their physical dimensions\cite{ziolkowski2006metamaterial} and enhance their performance and efficiency.\cite{ziolkowski2011metamaterial} Recently, the use of metasurfaces has widely been explored in the form of intelligent reflecting surfaces (IRSs)\cite{wu2019towards, sugiura2021joint, ino2023noncoherent, takimoto2025milli} or reconfigurable intelligent surfaces (RISs)\cite{zhang2018space, di2020smart, dai2020reconfigurable, gradoni1smart} to ensure sufficiently broad coverage, including non-line-of-site (NLOS) environments of high-frequency communications. However, although such metasurfaces are expected to play an important role in improving wireless communication environments, most metasurfaces are composed of nonlinear circuit components, biasing systems, and direct-current (DC) supplies, implying that the conventional RIS design requires an energy supply to activate the operation of RISs. Specifically, these metasurfaces have been utilized as components of active control systems, making it difficult to move RISs and thus limiting the use of unique electromagnetic properties in real-world communication environments.

\begin{figure}[tb!]
\includegraphics[width=\linewidth]{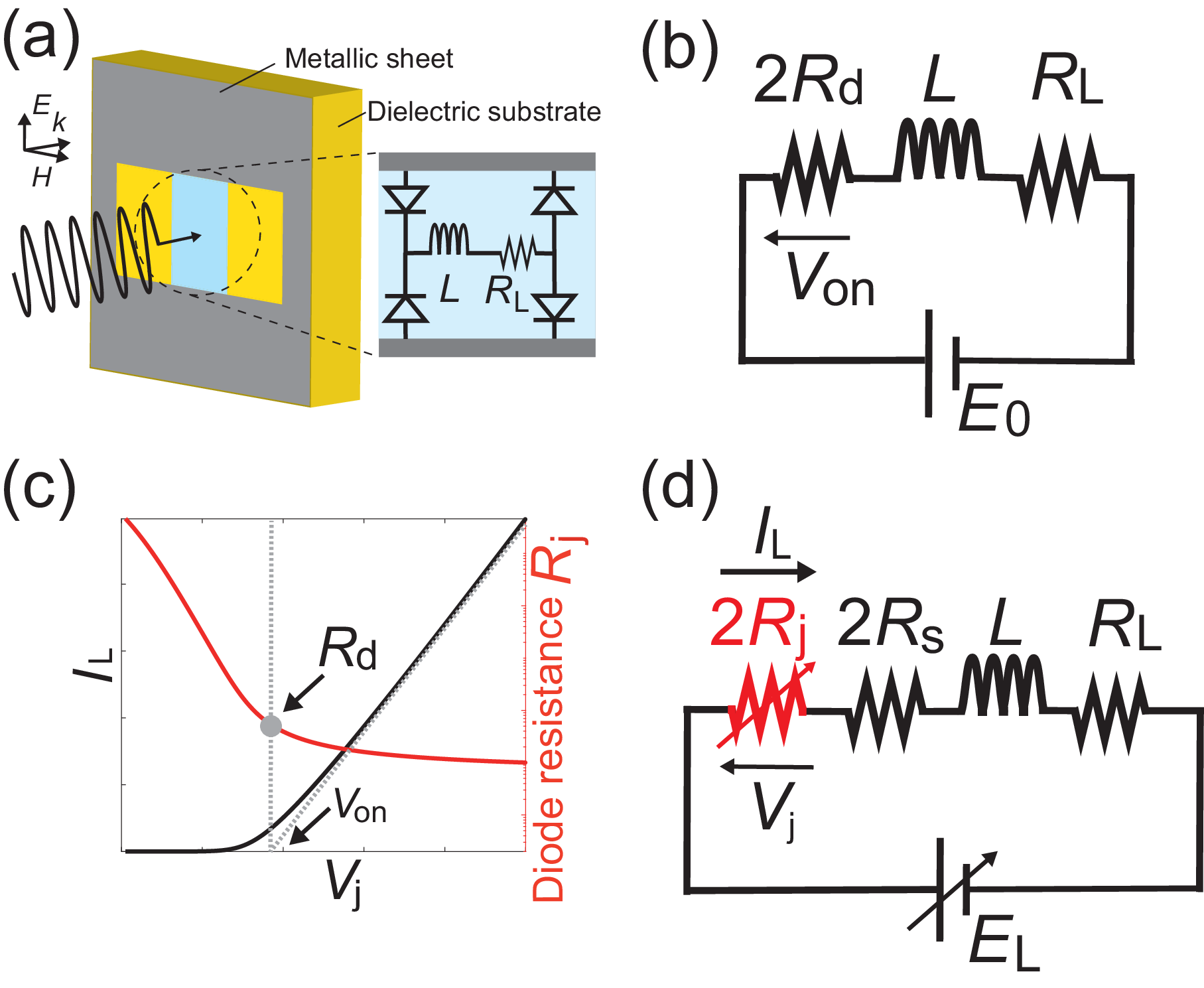}
\caption{\label{fig:1} Waveform-selective metasurface and simplified equivalent circuit models. (a) Periodic unit cell of an inductor-based waveform-selective metasurface. (b) Conventional simplified equivalent circuit model using $R_d$, which represents the effective resistive component of diodes only at their turn-on voltage $V_{on}$. (c) Current voltage characteristics of the diode. (d) Proposed simplified equivalent circuit model using $R_j$. Unlike the conventional model of (b), $R_j$ was used as a variable resistor whose resistance varied in accordance with the input power or the input voltage $E_L$ of (d).}
\end{figure}

In contrast to the active metasurface design, circuit-based metasurfaces were reported to control electromagnetic waves in a passive manner without external DC supplies, as shown in Fig.\ \ref{fig:1}a.\cite{wakatsuchi2013waveform, eleftheriades2014electronics, wakatsuchi2015waveformSciRep, vellucci2019waveform, f2020temporal, takeshita2024frequency} These metasurfaces, loaded with Schottky/PIN diodes to rectify incoming signals and exploit the transients of classic DC circuits, autonomously varied the scattering/absorbing profiles even at a fixed frequency in accordance with the incoming waveform (i.e., the incoming pulse width in the time domain). These waveform-selective metasurfaces have thus far been exploited for electromagnetic interference,\cite{wakatsuchi2019waveform} antenna and sensor design,\cite{ushikoshi2023pulse} signal processing,\cite{f2020temporal} and IRSs.\cite{fathnan2023unsynchronized} In particular, to fully exploit the advantages of waveform-selective metasurfaces in a large space of wireless communication environments, a simple design approach is needed for determining the optimal physical dimensions and circuit constants. With respect to the first issue (i.e., the optimal physical dimensions), equivalent circuit models have been reported by many research groups to understand metasurface behaviors,\cite{baena2005equivalent, ZhouCWeq, MyCWeqCircuitPaper} which can also be applied to estimate the operating frequencies of these time-varying metasurfaces. With respect to the second issue (i.e., the optimal circuit constants), the relationships between the waveform-selective response and the SPICE parameters of diodes were reported in the literature as design guidelines for rectifying diodes.\cite{imai2023design} Moreover, a simplified equivalent circuit approach was found to readily predict the time-varying electromagnetic responses of waveform-selective metasurfaces.\cite{aplEqCircuit4WSM, fathnan2022method} However, the conventional approach relies on a power-independent resistance, which fails to capture the nonlinear behaviors of the loaded diodes. This characteristic severely restricts the practical use of waveform-selective metasurfaces in diverse wireless communication environments, where input power levels dynamically fluctuate because of fading, shadowing, or varying distances between the transmitter and the metasurface. This discrepancy between the model and reality is not a minor inaccuracy; it represents a fundamental barrier to the predictive design and deployment of waveform-selective metasurfaces in the environments where they are most needed, such as dynamic wireless networks. The inability to accurately predict behaviors across different power levels makes robust system design nearly impossible. In this study, we directly address this challenge by introducing a simplified yet physically rigorous analytical model to predict the time-varying electromagnetic responses of waveform-selective metasurfaces at various input power levels. By adopting two mathematical solutions, namely, the Maclaurin series and the Wright omega function, our models show power- and time-dependent resistances and currents of diodes, as well as their scattering profiles, all of which are in close agreement with the numerically calculated results. 
Furthermore, we demonstrate that our approach is applicable to different waveform-selective metasurfaces, repeated pulsed signals, and no-signal-exposure periods; under these conditions, the signal intensity significantly changes. 
Furthermore, the proposed approach is utilized not only for resonant frequencies, where waveform-selective responses are most effectively exploited, but also for other frequencies near the resonant frequencies, which demonstrates the wide applicability of our proposed method.

\section{Theory and model}
Although the essential mechanisms of waveform-selective metasurfaces have been reported in many studies,\cite{wakatsuchi2019waveform, ushikoshi2023pulse, ozawa2025experimental} our metasurface was designed in the following manner. First, as shown in Fig.\ \ref{fig:1}a, our unit cell consisted of a dielectric substrate (1.5-mm-thick Rogers 3003) and a conducting surface including periodic rectangular slits (5 $\times$ 16 mm$^2$) with an 18-mm periodicity. Without any circuit elements, these metasurfaces, known as slit structures, resonated to strongly transmit incoming signals at their designed resonant frequencies.\cite{MunkBook, MTMbookEngheta, cwFiltering, yang2012reduction} However, this transmitting resonant mechanism could be controlled by loaded circuit elements. Specifically, a set of four Schottky diodes, which play the role of a diode bridge, was used to connect the conductor edges of each slit and control the degree of resonance intensity. Here, electric charges induced by the incident wave were fully rectified to convert the incoming frequency component to an infinite set of frequencies. Importantly, most energy was converted to a zero frequency, as seen from the Fourier expansion (considering the Fourier expansion of $|\sin|$ to derive each frequency term\cite{ushikoshi2023pulse}). Therefore, reactive and resistive circuit components were connected to the inside of the diode bridge to exploit the time-varying mechanism called the transients of classic DC circuits, which was efficiently coupled with the electromagnetic response of the slit structure. For instance, the intrinsic transmitting resonant mechanism of the slit structure was obtained during an initial period if an inductor ($L=$ 0.1 $\mu$H as a default value) was connected to a resistor ($R_L=$ 5.5 $\Omega$ as a default value) in series within the diode bridge as the inductor exhibited an electromotive force to prevent incoming electric charges.\cite{wakatsuchi2019waveform} However, this electromotive force gradually disappeared because of the zero-frequency component such that the intrinsic transmitting resonant mechanism of the slit structure weakened even at the same frequency. Thus, short pulses were expected to efficiently pass through the metasurface, whereas long pulses or continuous waves (CWs) were poorly transmitted even at the same frequency. Additional advanced pulse width-dependent responses could be designed depending on the loaded circuit configurations.\cite{wakatsuchi2015time}

In the literature, a waveform-selective response was effectively represented by the simplified equivalent circuit model drawn in Fig.\ \ref{fig:1}b. Here, the equivalent circuit model imposed two assumptions to facilitate simple yet effective predictions of the time-varying electromagnetic responses of waveform-selective metasurfaces.\cite{aplEqCircuit4WSM} Specifically, while the original incident signal was excited by an alternate current (AC) source, the voltage applied to the entire equivalent circuit model was represented by a DC source. This first assumption was valid for waveform-selective metasurfaces since, as explained above, most incident energy was converted to zero frequency because of the full wave rectification process of the diode bridge. Second, since deriving the rigorous time-varying voltage-dependent resistive component of the diodes was difficult, the diode bridge was represented by resistors with a fixed resistance value at their turn-on voltage in the literature. The second assumption was reasonable, as waveform-selective metasurfaces were mostly used when the incident signal intensity exceeded the turn-on voltage $V_{on}$ of the diodes, as shown in Fig.\ \ref{fig:1}c. However, this simplification could lead to a nonnegligible deviation if the incident power was largely varied, including in pulsed signal scenarios, which were addressed in this study.

\begin{figure}[tb!]
\includegraphics[width=\linewidth]{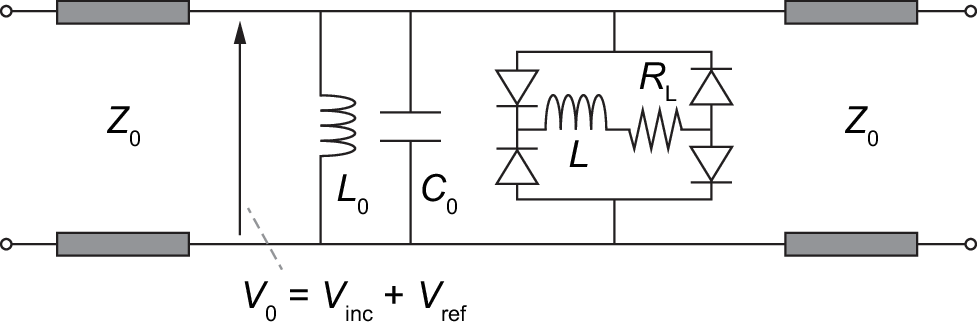}
\caption{\label{fig:1b} Transmission line model representing the waveform-selective metasurface shown in Fig.\ \ref{fig:1}a.}
\end{figure}

To overcome the restriction imposed by the power-independent resistance, we replaced fixed resistors $R_{d}$s with variable resistors $R_{j}$s. Additionally, our model included $R_s$, which represented the series resistance within the SPICE model of diodes. Importantly, the equivalent circuit model shown in Fig.\ \ref{fig:1}d had two unknown variables, namely, $R_j$ and the circuit current $I_L$, which implied that two equations were required to find $R_j$. To address this issue, the entire metasurface model was initially represented by the two-port network model shown in Fig.\ \ref{fig:1b}. In this figure, $Z_0$, $V_{inc}$, and $V_{ref}$ denote the wave impedance of vacuum and the incident and reflected voltages, respectively, while $V_0=V_{inc}+V_{ref}$. The entire capacitive and inductive components obtained by the conducting geometry and physical dimensions of our metasurface were approximated by $C_0$ and $L_0$. Here, $C_0$ and $L_0$ were connected to the diode bridge and its internal circuit components in parallel. At the resonant frequency, $C_0$ and $L_0$ had relatively large reactance magnitudes; thus, they could be ignored, as shown in Fig.\ \ref{fig:1}d, where $E_L=V_{inc}+V_{ref}$. Under these circumstances, using Kirchhoff's current and voltage laws, we derived the following circuit equation:
\begin{equation}
\left( 2R_j+ 2R_s+ R_L \right) I_L + \frac{dI_L}{dt}=E_L=2V_{inc}-Z_0I_L,
\label{eq1:kirch}
\end{equation}
where $t$ represents time. This equation could be rearranged to produce the following equation:
\begin{equation}
I_L=\frac{2V_{inc}}{2R_j+2R_s+R_L+Z_0}\left( 1-e^{-\frac{2R_j+2R_s+R_L+Z_0}{L}t} \right).
\label{eq2:kirch}
\end{equation}
To find $R_j$, we used the well-known relationship between the voltage and current of the Schottky diodes:
\begin{equation}
I_L=I_S e^{\frac{qV}{nKT}},
\label{eq3:diodeIV}
\end{equation}
where $I_S$, $q$, $V$, $n$, $K$, and $T$ denote the saturation current of the diodes, the charge of an electron, the voltage applied across the single diodes, the ideality factor (1.08 in this study), Boltzmann's constant, and the absolute temperature in Kelvin, respectively. Here, we omitted $-1$ within the exponential function. This value was negligible, as we assumed that the incident signal intensity was sufficiently large. Eq.\ (\ref{eq3:diodeIV}) could be rearranged as follows:
\begin{equation}
V_j=\frac{nkT}{q}\ln{\frac{I_L}{I_S}},
\label{eq3b:diodeIV}
\end{equation}
where $V_j$ represents the voltage across $R_j$. Thus, $R_j$ became
\begin{equation}
R_j=\frac{\frac{nKT}{q}\ln{\frac{I_L}{I_S}}}{I_L}.
\label{eq4:diodeIV}
\end{equation}
By substituting Eq.\ (\ref{eq4:diodeIV}) into Eq.\ (\ref{eq2:kirch}), the following equation could be derived:
\begin{widetext}
\begin{eqnarray}
\dfrac{2V_{inc}R_j}{2R_j+2R_s+R_L+Z_0}\left( 1-e^{-\frac{2R_j+2R_s+R_L+Z_0}{L}t}\right)=\frac{nKT}{q}\ln{\left( \frac{2V_{inc}}{I_S\left( 2R_j+2R_s+R_L+Z_0\right)}\left( 1-e^{-\frac{2R_j+2R_s+R_L+Z_0}{L}t}\right) \right)}.
\label{eq5:rigorous}
\end{eqnarray}    
\end{widetext}
Clearly, $R_j$ could not be easily calculated in this equation, as $R_j$ appeared not only in exponential functions but also in a natural logarithmic function.

To find $R_j$, we therefore applied two mathematical approaches. First, we used the Maclaurin series and a simplified approximation based on a rational function (Appendix A) to replace the exponential functions of Eq.\ (\ref{eq5:rigorous}) with
\begin{eqnarray}
1-e^{-\frac{2R_j+2R_s+R_L+Z_0}{L}t} \approx \frac{1}{\frac{L}{\left( 2R_j+2R_s+R_L+Z_0 \right) t}+1}.
\label{eq6:Laurent}
\end{eqnarray}
This approximation replaced the exponential term, which represents the transient response of the RL circuit,\cite{aplEqCircuit4WSM} with a hyperbolic saturation curve (Eq. (25)). As detailed in Appendix A, this rational function effectively captured the essential behavior of the transient response, both in the initial phase (small $t$) and the steady state (large $t$), thereby enabling an analytical solution in the subsequent step using an additional mathematical function (the validity of this approximation was confirmed by the close agreement with the numerical results shown in Section \ref{sec:results}). By plugging Eq.\ (\ref{eq6:Laurent}) into Eq.\ (\ref{eq5:rigorous}), we obtained
\begin{widetext}
\begin{eqnarray}
\frac{qV_{inc}\left( L+t(2R_s+R_L+Z_0) \right)}{nKT\left( L+t\left( 2R_j + 2R_s+R_L+Z_0\right) \right)}+\ln{\frac{qV_{inc}\left( L+t(2R_s+R_L+Z_0) \right)}{nKT\left( L+t\left( 2R_j + 2R_s+R_L+Z_0\right) \right)}}=\frac{qV_{inc}}{nKT}-\ln{\frac{2ntKT}{qI_S\left( L+t(2R_s+R_L+Z_0)\right)}}.
\label{eq7:Laurent}
\end{eqnarray}
\end{widetext}
Second, we used the Wright omega function $\omega$, which is a Lambert $W$ function suited for solving a natural logarithmic function. Specifically, an unknown variable $X$ in the expression of $X+\ln{X}=Y$ could be solved via $X=\omega (Y)$. Thus, we assumed the following:
\begin{eqnarray}
X=\frac{qV_{inc}\left( L+t(2R_s+R_L+Z_0) \right)}{nKT\left( L+t\left( 2R_j + 2R_s+R_L+Z_0\right) \right)},
\label{eq8:WOFx}
\end{eqnarray}
\begin{eqnarray}
Y=\frac{qV_{inc}}{nKT}-\ln{\frac{2ntKT}{qI_S\left( L+t(2R_s+R_L+Z_0)\right)}}.
\label{eq8:WOFy}
\end{eqnarray}
By using these equations, $X$ (including only one $R_j$) was calculated as follows:
\begin{eqnarray}
X = \omega \left( \frac{qV_{inc}}{nKT}-\ln{\frac{2ntKT}{qI_S\left( L+t(2R_s+R_L+Z_0)\right)}} \right),
\label{eq9:WOF}
\end{eqnarray}
where $\omega$ represents the Wright omega function mentioned above. Therefore, with Eq.\ (\ref{eq8:WOFx}), $R_j$ became 
\begin{widetext}
\begin{eqnarray}
R_j=\frac{1}{2}\left( \frac{1}{t}\left( \frac{qV_{inc}(L+t(2R_s+R_L+Z_0))}{nKTX}-L\right) -2R_s-R_L-Z_0\right).
\label{eq10:Rj}
\end{eqnarray}
\end{widetext}

Next, using the $R_j$ value obtained in Eq.\ (\ref{eq10:Rj}), we estimated the transmittance of our metasurface. Here, we adopted the ABCD matrix approach to approximate the wave propagation through the metasurface:
\begin{eqnarray}
M_{ms}&=&
\begin{bmatrix}
1 & 0 \\
1/Z_c & 1 \\
\end{bmatrix}
\begin{bmatrix}
\cos{(\beta _sh)} & jZ_s\sin{(\beta _sh)} \\
j\sin{(\beta _sh)}/Z_s & \cos{(\beta _sh)} \\
\end{bmatrix}
\nonumber\\
&=&
\begin{bmatrix}
A & B \\
C & D \\
\end{bmatrix}
,\label{eq11:ABCD}
\end{eqnarray}
where $M_{ms}$ is the ABCD matrix of the metasurface. Also, $\beta_s$, $h$, and $Z_s$ are the wavenumber, thickness, and wave impedance of the metasurface substrate, respectively, and $Z_c$ represents the circuit impedance of the metasurface. Since we mostly considered the resonant state of the metasurface, the pair of $L_0$ and $C_0$ in Fig.\ \ref{fig:1b} behaved as an open circuit. Under these circumstances, the incident voltage was applied across the diode bridge of Fig.\ \ref{fig:1b} so that $Z_c$ was obtained from the relationship between the voltage and current of the diode bridge, i.e.,
\begin{eqnarray}
Z_c&=&\frac{V_0}{I_L}=\frac{2V_{inc}-Z_0I_{L}}{I_L}\nonumber\\
&=&\frac{2R_j+2R_s+R_L+Z_0}{1-e^{-\frac{2R_j+2R_s+R_L+Z_0}{L}t}}-Z_0.
\label{eq12:Zc}
\end{eqnarray}
Therefore, by substituting Eq.\ (\ref{eq12:Zc}) into Eq.\ (\ref{eq11:ABCD}) with Eq.\ (\ref{eq10:Rj}), the transmission coefficient $S_{21}$ of our metasurface was calculated from the following:
\begin{eqnarray}
S_{21}=\frac{2}{A+B/Z_0+CZ_0+D}.
\label{eq13:T}
\end{eqnarray}
$S_{21}$ was then squared (i.e., $S_{21}^2$) to estimate the transmittance of each metasurface.

\section{Results}
\label{sec:results}
\subsection{Fundamental results using inductor-based waveform-selective metasurfaces}

\begin{figure}[tb!]
\includegraphics[width=\linewidth]{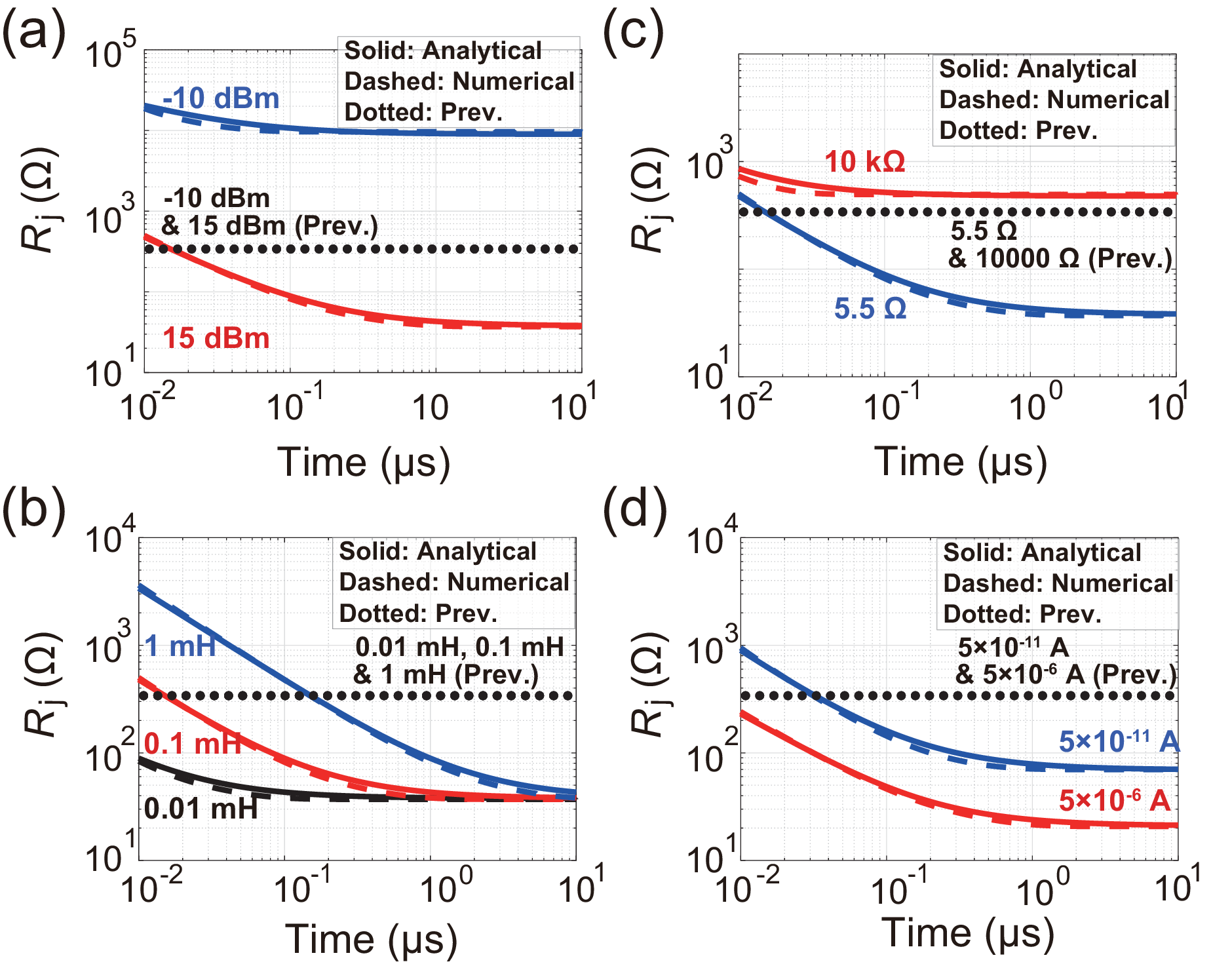}
\caption{\label{fig:2} Estimated power- and time-dependent $R_j$. The results with various (a) input powers, (b) $L$s, (c) $R_L$s, and (d) $I_s$s.}
\end{figure}

The abovementioned analytical solution was evaluated by using three indices, specifically, the resistances of the diodes, their currents, and the electromagnetic responses of inductor-based waveform-selective metasurfaces. As the first evaluation, we estimated the power-dependent time-varying resistive components of the diodes, namely, $R_j$. The analytically derived $R_j$ values were compared with the results obtained by numerical simulations based on a cosimulation method available in ANSYS Electronics Desktop (2022R2).\cite{ushikoshi2023pulse, takeshita2024frequency} The numerical resistance value was obtained by dividing the voltage across one of the diodes in the metasurface by the diode current. A comparison of the results is shown in Fig.\ \ref{fig:2}, where the input power, $L$, $R$, and $I_S$ were changed. Here, we set the default values of the input power, $L$, $R$, and $I_S$ to 15 dBm, 0.1 mH, 5.5 $\Omega$, and 5.0 $\times$ 10$^{-6}$ A, respectively. As a result, our equivalent circuit model using the variable $R_j$ effectively predicted the numerically derived values. As reference data, we plotted the constant resistance value used in the conventional method, namely, $R_d$, as shown in Fig.\ \ref{fig:1}b and Fig.\ \ref{fig:1}c.\cite{aplEqCircuit4WSM} Clearly, $R_d$ remained the same despite the changes in the input power, $L$, $R$, and $I_S$ in Fig.\ \ref{fig:2}.

\begin{figure}[tb!]
\includegraphics[width=\linewidth]{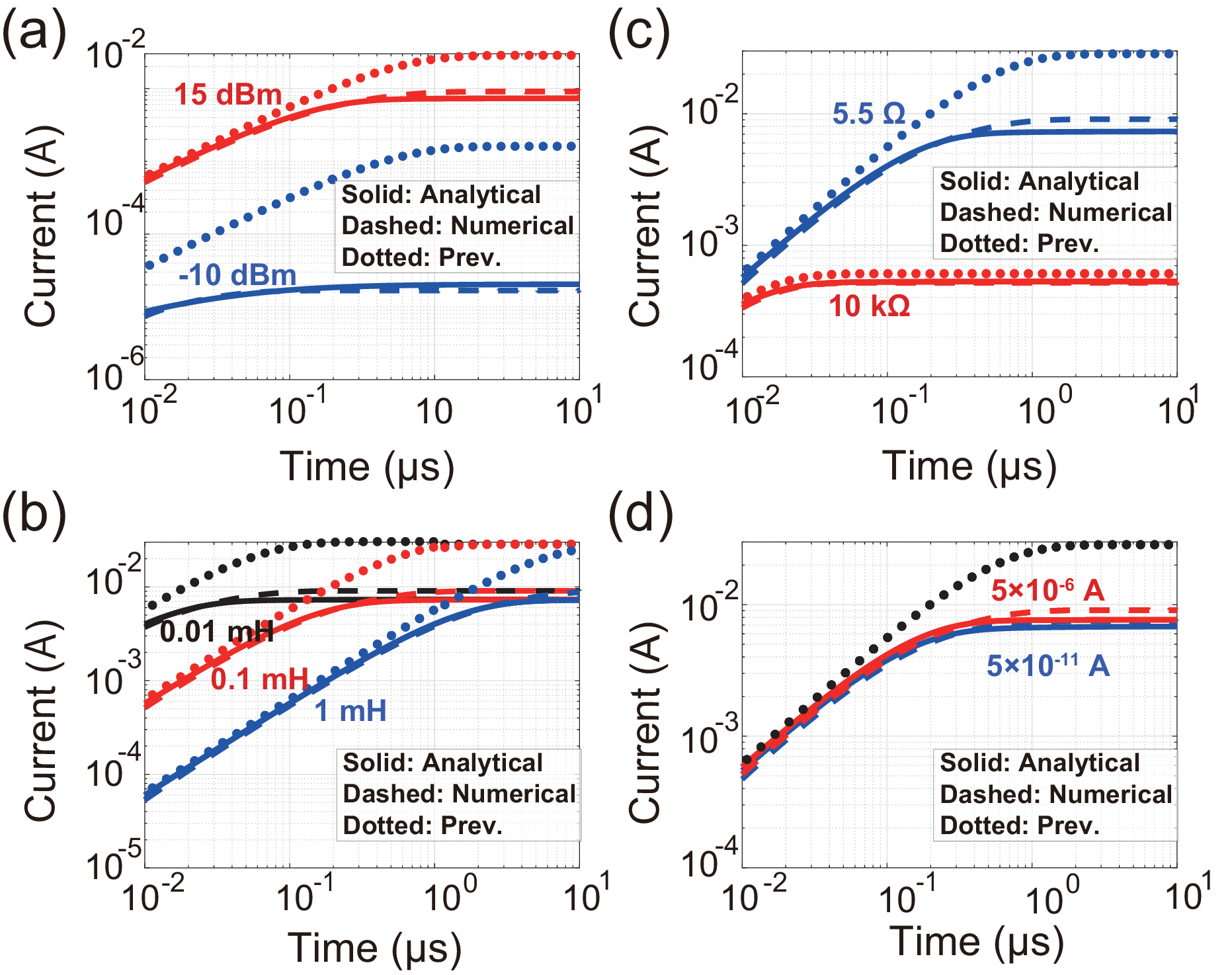}
\caption{\label{fig:3} Diode current. The results with various (a) input powers, (b) $L$s, (c) $R_L$s, and (d) $I_s$s.}
\end{figure}

Second, by using Eq.\ (\ref{eq2:kirch}) with the $R_j$ values derived in Fig.\ \ref{fig:2}, the diode current (i.e., another unknown variable in Fig.\ \ref{fig:1}d) was evaluated in Fig.\ \ref{fig:3}. In this figure, we evaluated the influences of the input power, $L$, $R$, and $I_S$, as demonstrated in Fig.\ \ref{fig:2}. As a result, as shown in Fig.\ \ref{fig:3}, when the input power was set to a sufficiently large value to turn on the diodes, such as 15 dBm, the conventional approach using the constant $R_d$ value showed relatively close agreement with the numerically derived diode current. However, a large deviation appeared when the input power was reduced (see the result of -10 dBm in the blue dotted curve in Fig.\ \ref{fig:3}a). Physically, at a low input power, the voltage across the diodes was low, resulting in a high diode resistance $R_{j}$. The conventional approach, which used a constant $R_{d}$ fixed near the turn-on voltage, significantly underestimated the actual resistance in this regime, leading to the observed deviation. In contrast, our proposed approach accurately reproduced the numerically derived current across a range of input power levels. This accuracy was achieved because the model effectively captured the power-dependent and time-varying resistance of the diode (Fig.\ \ref{fig:2}).

\begin{figure}[tb!]
\includegraphics[width=\linewidth]{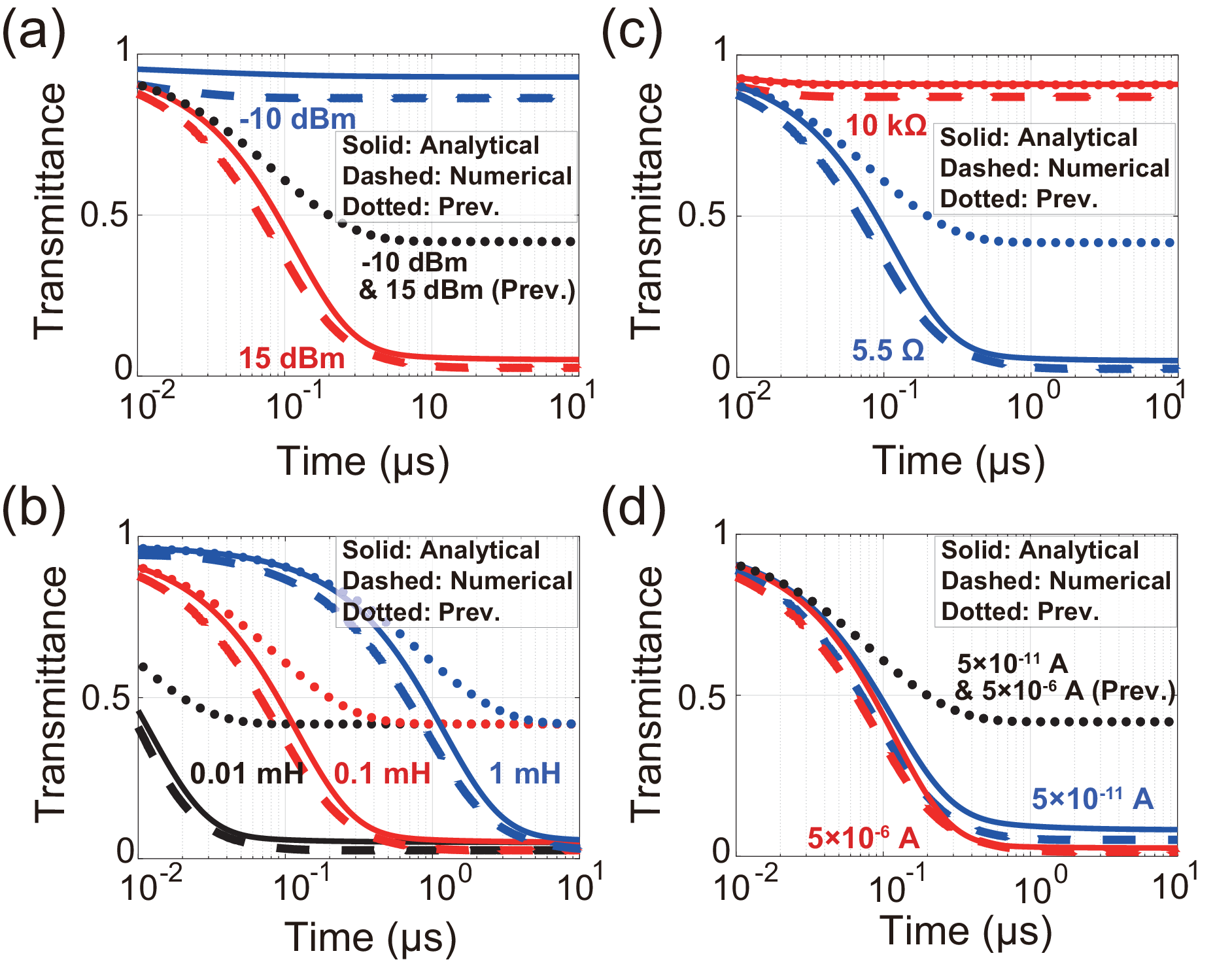}
\caption{\label{fig:4} Transmittance. The results with various (a) input powers, (b) $L$s, (c) $R_L$s, and (d) $I_s$s.}
\end{figure}

Third, we estimated the transmittance of the inductor-based waveform-selective metasurface in Fig.\ \ref{fig:4} with the numerically derived results and the analytical results based on the conventional method using the constant resistive value (i.e., $R_d$). As a result, compared with that of the conventional approach, the transmittance of the proposed method closely matched that of the numerical method, as shown in Fig.\ \ref{fig:4}. All these results could be attributed to the estimation accuracy of the diode resistance, as explained in Figs.\ \ref{fig:2} and \ref{fig:3}.

The close agreement between our analytical model and the numerical results shown in Fig.\ \ref{fig:4} was a direct consequence of the ability of the model to accurately predict the power- and time-dependent diode resistance $R_j$, as established in Fig.\ \ref{fig:2}. By correctly capturing this fundamental parameter, the model subsequently predicted the circuit current with high fidelity (Fig.\ \ref{fig:3}), which in turn determined the overall metasurface impedance and final transmittance. In contrast, the large discrepancy in the transmittance prediction of the conventional method could be traced back to its fundamental failure to represent the true behavior of the diode, particularly at lower power levels (e.g., -10 dBm), with a fixed resistance value.

\begin{figure}[tb!]
\includegraphics[width=\linewidth]{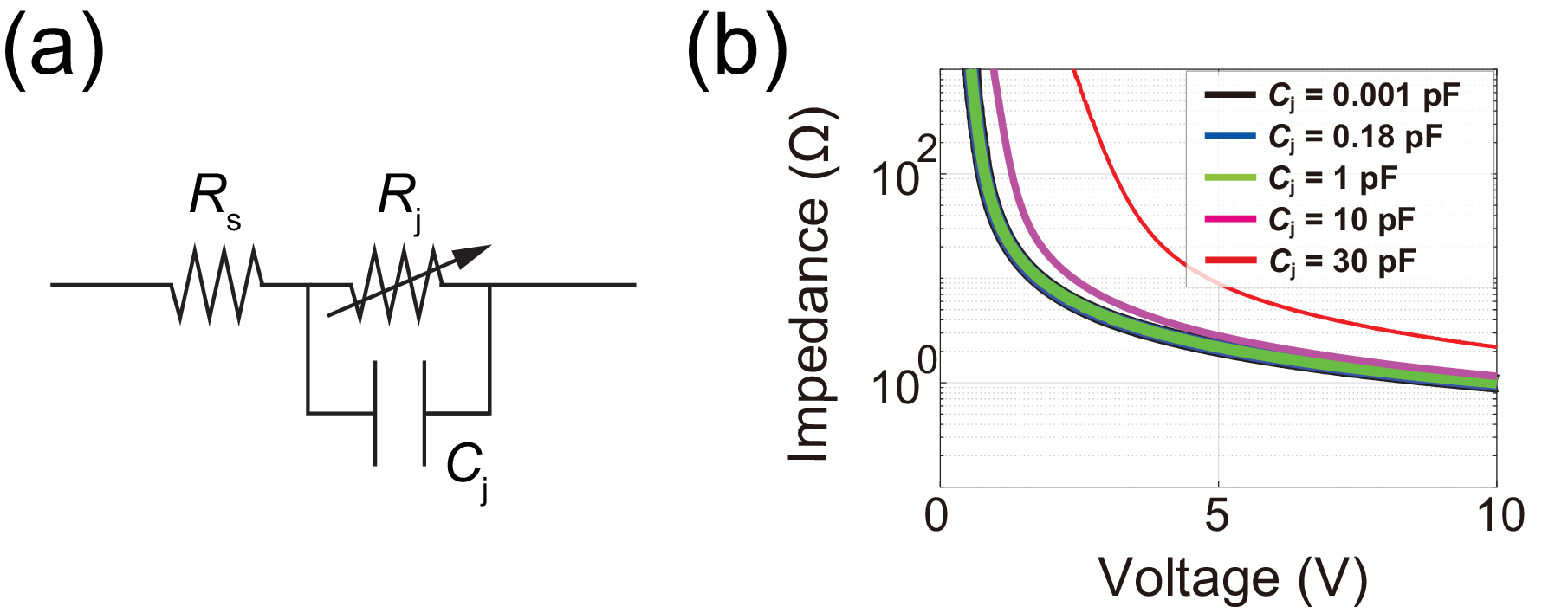}
\caption{\label{fig:5} Influence of the parasitic capacitor $C_j$. (a) Simplified diode representation. (b) Calculation results of the diode impedance.}
\end{figure}

To further analyze the performance and the limitations of the proposed method, we evaluated the influences of the parasitic (junction) capacitances $C_j$ of the diodes in Fig.\ \ref{fig:5}. This analysis was important for assessing the performance of our proposed approach, as it omitted the minor influence occurring from $C_j$ in finding $R_j$, while $R_j$ was often connected to $C_j$ in parallel, as shown in Fig.\ \ref{fig:5}a. Although ordinary diodes contained $C_j$, our simplification became reasonable only if the $C_j$ value was limited. For instance, our diode model (Avago, HSMS-286x series) had a junction capacitor of $C_j=$ 0.18 pF. This value largely varied in Fig.\ \ref{fig:5}b, which showed how the impedance changed depending on the input voltage. As a result, the impedance remained unchanged and almost the same as $R_j$ when $C_j<$ 10 pF, which was significantly greater than the junction capacitance of our diode model (i.e., 0.18 pF). Thus, these results supported the concept that the proposed method was valid for an ordinary range of $C_j$.

\begin{figure}[tb!]
\includegraphics[width=\linewidth]{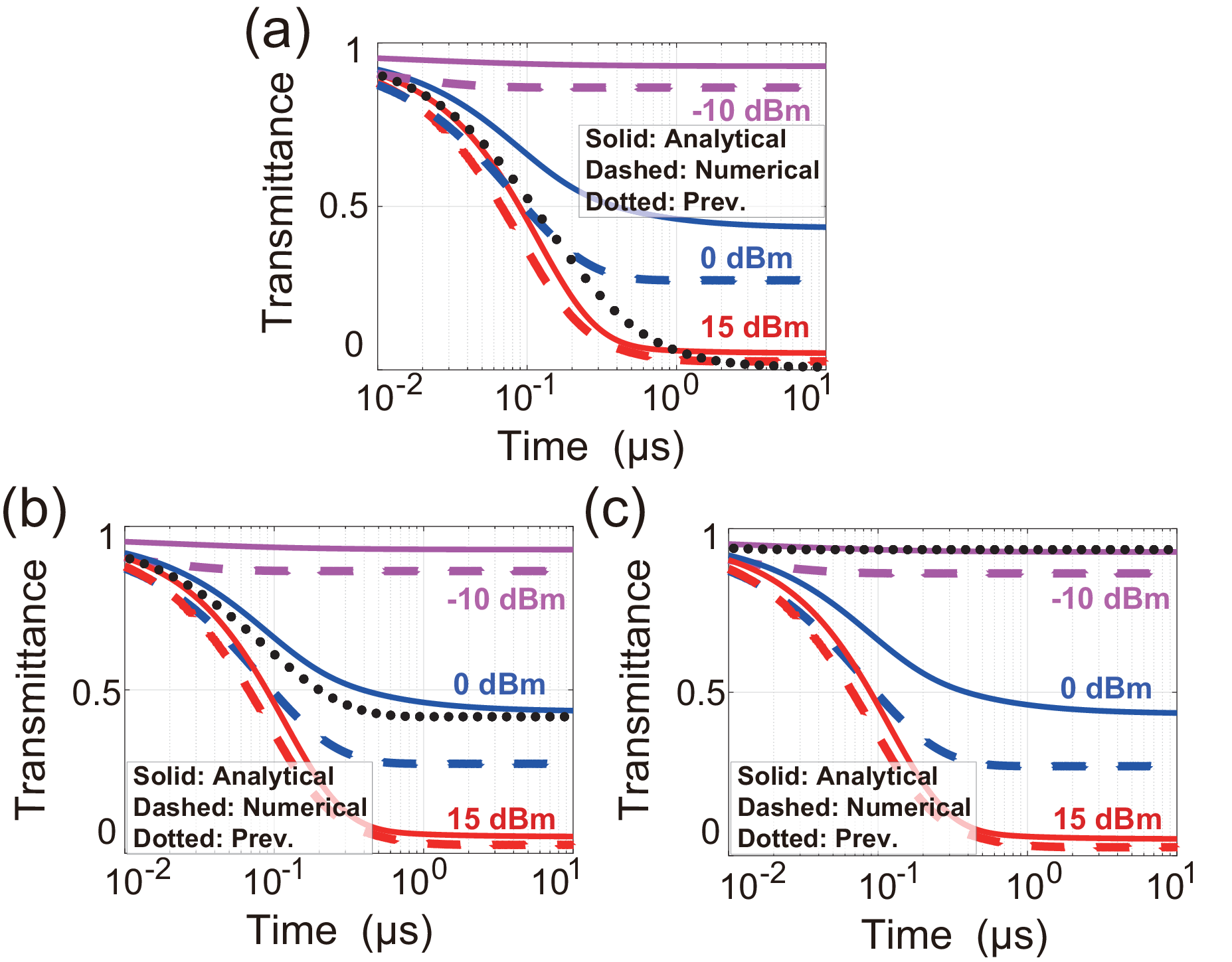}
\caption{\label{fig:6} Conventional approach using various $R_d$ values. Comparison of results using $R_d=$ (a) 10 $\Omega$, (b) 340 $\Omega$, and (c) 10 k$\Omega$.}
\end{figure}

We analyzed the relationships between the numerical results and the analytical results based on our proposed method and the conventional method. As explained above, the conventional method used a constant diode resistance value at the turn-on voltage to simplify the diode behavior as a single value. In fact, this value could be changed to any other arbitrary number to reduce the deviation from the numerically derived results. As shown in Fig.\ \ref{fig:6}, by varying $R_d$, the numerical simulation results could be closer to the results of the conventional analytical approach than those of the proposed approach. However, these favorable agreements were obtained coincidentally by adjusting the $R_d$ value to achieve a better outcome, unlike our proposed approach, which showed close agreement without these artificial adjustments. Notably, an important role of simplified equivalent circuit models was its ability to readily predict the responses of waveform-selective metasurfaces without requiring numerical simulations to facilitate the entire design of waveform-selective metasurfaces. Thus, avoiding such a repetition of numerical simulations was considered more ideal.

\subsection{Extended concept and results}

\begin{figure}[tb!]
\includegraphics[width=\linewidth]{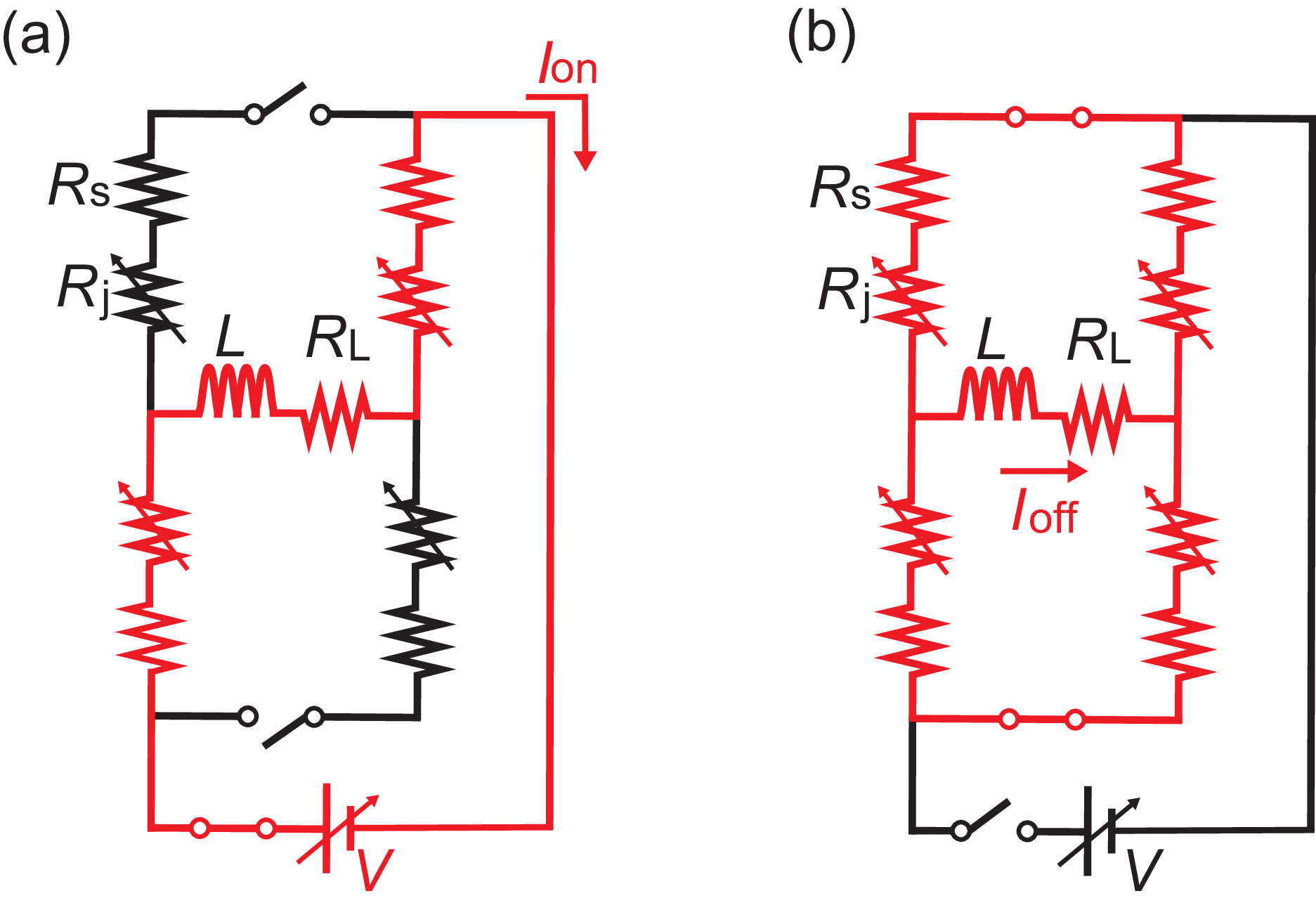}
\caption{\label{fig:7} Simplified equivalent circuit models representing time-domain responses to repeated pulses: (a) on state (under pulse illumination) and (b) off state (between pulses).}
\end{figure}

Our fundamental approach was explained above, while in this subsection, we explained how our method could be utilized for more advanced application scenarios, including repeated pulse scenarios, different waveform-selective metasurfaces, and other frequencies. First, the above results considered only a simple CW scenario, whereas realistic communication environments included no-signal durations. This scenario was considered using the simplified equivalent circuit models drawn in Fig.\ \ref{fig:7}. Here, we evaluated our metasurface (or our model) with repeated pulses oscillating at a constant frequency. Our metasurface was represented by two different states, an on state and an off state, each of which was assumed to operate with and without an incident pulse, respectively. Essentially, the former equivalent circuit model shown in Fig.\ \ref{fig:7}a was the same as the one adopted in Fig.\ \ref{fig:2} to Fig.\ \ref{fig:4} (i.e., the model shown in Fig.\ \ref{fig:1}d). Therefore, $R_j$ was calculated in the same manner as before, except for a few changes to properly take over the voltage of the previous state as an initial condition:
\begin{widetext}
\begin{eqnarray}
R_j=\frac{1}{2}\left( \frac{1}{t}\left( \frac{q(2tV_{inc}+Li')(L+t(2R_s+R_L+Z_0))}{2ntKTX}-L\right) -2R_s-R_L-Z_0\right),
\label{eq10xx:ON}
\end{eqnarray}
\end{widetext}
where $i'$ denotes the initial value of the $I_{on}$ current in Fig.\ \ref{fig:7}a. Additionally, $X$ was determined by
\begin{widetext}
\begin{eqnarray}
X = \omega \left( \frac{q(2tV_{inc}+Li')}{2ntKT}-\ln{\frac{2ntKT}{qI_S\left( L+t(2R_s+R_L+Z_0)\right)}} \right).
\label{eq10xx:ON2}
\end{eqnarray}
\end{widetext}
These equations were used with Eqs.\ (\ref{eq11:ABCD}) to (\ref{eq13:T}) to estimate the transmittance during pulsed signals.
The equivalent circuit model was changed between the pulsed signals, as shown in Fig.\ \ref{fig:7}b. By adopting the same approach as above, we derived $R_j$ as follows:
\begin{widetext}
\begin{eqnarray}
R_j=\frac{1}{t}\left( \frac{qLi''(L+t(2R_s+R_L+Z_0))}{ntKTX}-L\right) -R_s-R_L,
\label{eq11xx:OFF}
\end{eqnarray}
\end{widetext}
where $i''$ denotes the initial value of the $I_{off}$ current in Fig.\ \ref{fig:7}b. Here, $X$ was changed to
\begin{widetext}
\begin{eqnarray}
X = \omega \left( \frac{qLi''}{ntKT}-\ln{\frac{ntKT}{qI_S\left( L+t(R_s+R_L+Z_0)\right)}} \right).
\label{eq11xx:OFF2}
\end{eqnarray}
\end{widetext}
Similarly, these equations were used with Eqs.\ (\ref{eq11:ABCD}) to (\ref{eq13:T}) to predict the transmittance between pulsed signals.

\begin{figure}[tb!]
\includegraphics[width=\linewidth]{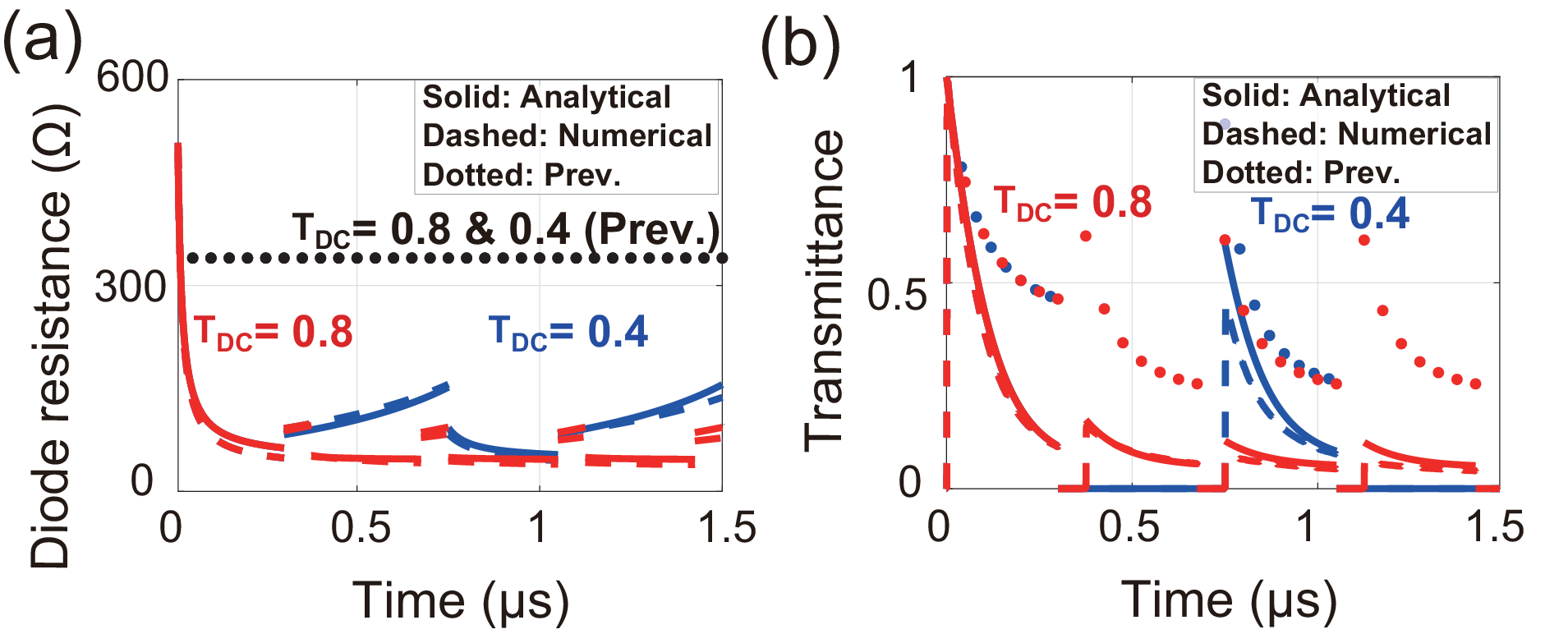}
\caption{\label{fig:8} Estimation results for repeated pulses. The results for (a) $R_j$ and (b) transmittance.}
\end{figure}

By using Eqs.\ (\ref{eq10xx:ON}) to (\ref{eq11xx:OFF2}), $R_j$ was obtained as shown in Fig.\ \ref{fig:8}a, where the incident pulse width was set to 300 ns, while the duty cycle $T_{DC}$ was set to either 0.4 or 0.8. We set the input power, $L$, $R_L$, and $I_S$ to 15 dBm, 0.1 mH, 5.5 $\Omega$, and 5.0 $\times$ 10$^{-8}$ A, respectively. According to Fig.\ \ref{fig:8}a, the analytically derived $R_j$ closely agreed with the numerically calculated value even if the pulse illumination stopped. Therefore, as shown in Figs.\ \ref{fig:2} to \ref{fig:4}, the use of the analytical $R_j$ value led to close matching with the numerically calculated transmittances, which was plotted in Fig.\ \ref{fig:8}b. Notably, the previous method using a constant $R_d$ value resulted in significantly large deviations in Fig.\ \ref{fig:8}b, as the corresponding diode resistance was largely different from the numerical value of the diode resistance (Fig.\ \ref{fig:8}a).

\begin{figure}[tb!]
\includegraphics[width=\linewidth]{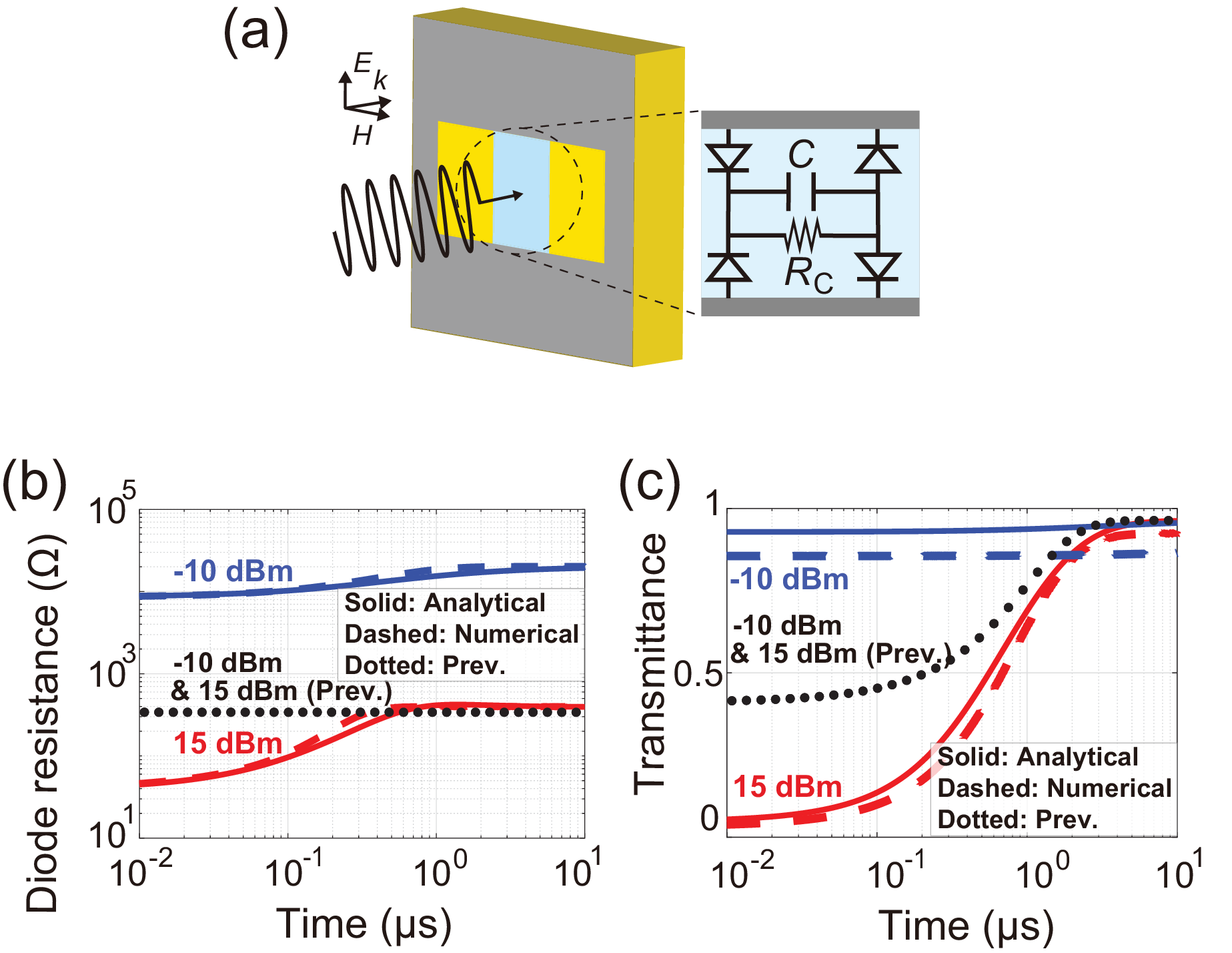}
\caption{\label{fig:9} Model for a capacitor-based waveform-selective metasurface and results. (a) Periodic unit cell of the capacitor-based waveform-selective metasurface. The results for (b) $R_j$ and (c) transmittance.}
\end{figure}

Second, the proposed approach was applied to a different type of structure, specifically the capacitor-based waveform-selective metasurface shown in Fig.\ \ref{fig:9}a. This structure is known to be able to more strongly transmit CWs than short pulses at the same frequency.\cite{wakatsuchi2015waveformSciRep, wakatsuchi2019waveform} By using the same procedures as those explained above, $R_j$ was derived as follows:
\begin{widetext}
\begin{eqnarray}
R_j=\frac{1}{2}\left( \frac{qV_{inc}(tR_C+(2R_s+Z_0)(CR_C+t))}{nKTX(CR_C+t)}-\frac{tR_C}{CR_C+t} -2R_s-Z_0\right),
\label{eq20:Rj}
\end{eqnarray}
\end{widetext}
where $X$ was changed to
\begin{eqnarray}
X = \omega \left( \frac{qV_{inc}}{nKT}-\ln{\frac{2nKT(CR_C+t)}{qI_S\left( tR_C+(2R_s+Z_0)(CR_C+t)\right)}} \right).
\label{eq21:WOF}
\end{eqnarray}
Here, $\omega$ represents the Wright omega function mentioned above. Notably, for the capacitor-based waveform-selective metasurface, $Z_C$ (Eq.\ (\ref{eq12:Zc})) had to be updated because the equivalent circuit was different from the one drawn in Fig.\ \ref{fig:1b}. In addition, the same procedure could be adopted to estimate $R_j$ and the transmittance of the capacitor-based waveform-selective metasurface. For example, Fig. \ref{fig:9}b and Fig. \ref{fig:9}c show $R_j$ and the transmittance while using the 15-dBm input power, $C=0.1$ nF, $R_C=10$ k$\Omega$, and $I_S=$ 5.0 $\times$ 10$^{-8}$ A (see $C$ and $R_C$ in Fig.\ \ref{fig:9}a). According to these results, compared with the conventional approach, the proposed method effectively predicted the power- and time-dependent characteristics of $R_j$ and the transmittance of the capacitor-based waveform-selective metasurface.

\begin{figure}[tb!]
\includegraphics[width=\linewidth]{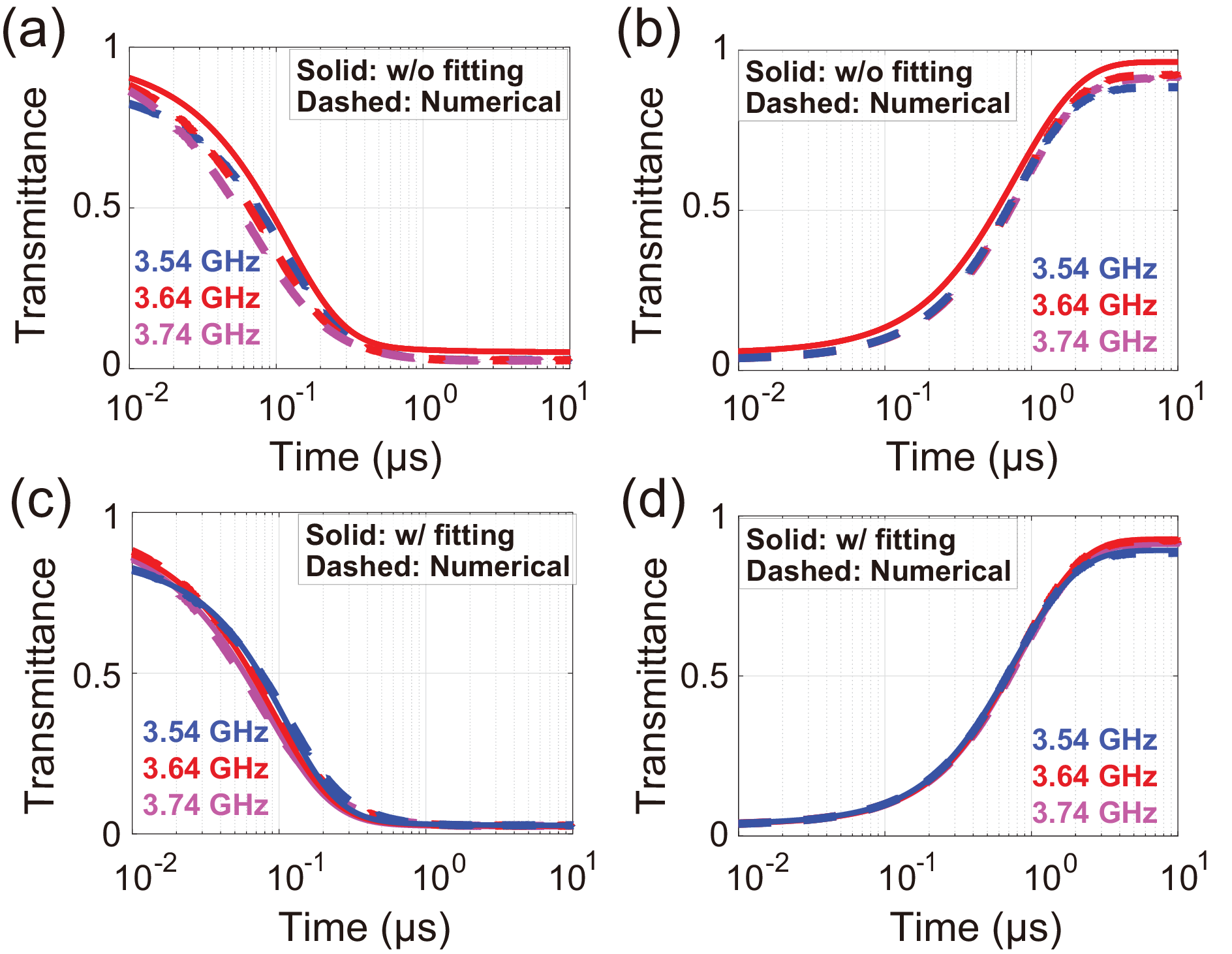}
\caption{\label{fig:10} Transmittance estimation using the fitting parameters of Table \ref{tab:1} for other frequencies. The results without the fitting parameters for (a) capacitor- and (b) inductor-based waveform-selective metasurfaces. The results with the fitting parameters for (c) the capacitor- and (d) inductor-based waveform-selective metasurfaces. 3.64 GHz represented their resonant frequency, whereas 3.54 and 3.74 GHz were slightly off the resonant frequency (but still close enough to show time-varying responses).}
\end{figure}

Finally, we explored the possibility of applying the proposed method to nonresonant frequencies, as our method has thus far been evaluated at resonant frequencies. On the one hand, the time-varying behaviors of waveform-selective metasurfaces were effectively obtained at resonance frequencies, while this response could possibly be obtained at other frequencies if the frequencies were not too far from the resonance points. On the other hand, our approach was made possible by assuming that the pair of $L_0$ and $C_0$ in Fig.\ \ref{fig:1b} behaved as an open circuit, indicating that the incident voltage (or the incident power) was fully applied to the diode bridge in Fig. \ref{fig:1b}. This assumption was no longer valid for nonresonant frequencies, as the incident energy was not fully coupled to the diode bridge. To address this issue and expand the applicability of the proposed method at nonresonant frequencies that were still close to the center of the resonance frequencies, we introduced fitting parameters.\cite{fathnan2022method} In this approach, we assumed that the entire shunt admittance $Y_{es}$ of our metasurface consisted of
\begin{eqnarray}
Y_{es}=Y_{freq}+Y_{time},
\end{eqnarray}
where $Y_{freq}$ and $Y_{time}$ represent the shunt impedances varying in the frequency and time domains, respectively. As adopted for ordinary metasurfaces,\cite{MunkBook, MTMbookEngheta, fathnan2022method} $Y_{freq}$ was determined by the physical geometry and material properties (e.g., conductors and substrate as well as their physical geometries), whereas $Y_{time}$ was related to the loaded circuit components. Since our metasurface could be represented by an inductor $L_0$ connected to capacitor $C_0$ in parallel, as shown in Fig.\ \ref{fig:1b}, $Y_{freq}$ was obtained by\cite{fathnan2022method}
\begin{eqnarray}
Y_{freq}=j\left(2\pi fC_0-\frac{1}{2\pi fL_0}\right) + \dot{a},
\end{eqnarray}
where $f$ is the frequency and $\dot{a}=a'+ja''$, and $a'$ and $a''$ represent the real and imaginary parts of $\dot{a}$, respectively. Additionally, $Y_{time}$ was calculated by\cite{fathnan2022method}
\begin{eqnarray}
Y_{time}=\dot{b}\frac{i_{inc}(t)}{V_{inc}},
\end{eqnarray}
where $i_{inc}$ denotes the incident current of the equivalent circuit model of the metasurface. Additionally, similar to $\dot{a}$, $\dot{b}$ was $\dot{b}=b'+jb''$, with $b'$ and $b''$ representing the real and imaginary parts of $\dot{b}$, respectively. Under these circumstances, using the fitting parameters of Table \ref{tab:1}, we calculated the transmittances of waveform-selective metasurfaces at 3.64 (the center of their resonant frequency), 3.54, and 3.74 GHz, namely, two additional frequencies near the resonant frequency, as shown in Fig.\ \ref{fig:10}. Here, we set the input power, $L$, $R_L$, $C$, $R_C$, and $I_S$ to 15 dBm, 0.1 mH, 5.5 $\Omega$, 1 nF, 10 k$\Omega$, and 5.0 $\times$ 10$^{-8}$ A, respectively. As a result, Fig.\ \ref{fig:10}a and Fig.\ \ref{fig:10}b show that without the fitting parameters, our equivalent circuit approach was close to the numerically derived results but with slight deviations (see that all three solid curves were the same in each graph). However, by introducing the fitting parameters in Fig.\ \ref{fig:10}c and Fig.\ \ref{fig:10}d, closer agreement was obtained between our approach and the numerical results at each frequency, including 3.54 and 3.74 GHz (i.e., the two frequencies off the resonant frequency). These results indicated that our equivalent circuit approach improved the estimation accuracy by using fitting parameters to adjust the coupling between the incident wave and the metasurface circuit.

\begin{table*}[tb!]
\caption{\label{tab:1} Fitting parameters adopted for Fig.\ \ref{fig:10}.}
\centering
\renewcommand{\arraystretch}{1.2} 
\begin{tabular}{
|>{\centering\arraybackslash}p{1.8cm}||
>{\centering\arraybackslash}p{1.8cm}|
>{\centering\arraybackslash}p{1.8cm}|
>{\centering\arraybackslash}p{1.8cm}|
>{\centering\arraybackslash}p{1.8cm}|
>{\centering\arraybackslash}p{1.8cm}|
>{\centering\arraybackslash}p{1.8cm}|
}
\hline
\multirow{3}{*}{} 
& \multicolumn{3}{c|}{Inductor-based} 
& \multicolumn{3}{c|}{Capacitor-based} \\ 
& \multicolumn{3}{c|}{waveform-selective metasurfaces} 
& \multicolumn{3}{c|}{waveform-selective metasurfaces} \\ 
\cline{2-7}
& 3.54 GHz & 3.64 GHz & 3.74 GHz & 3.54 GHz & 3.64 GHz & 3.74 GHz \\ 
\hline\hline
$a'$  & 1.15 & 1.15 & 1.15 & 0.95 & 0.95 & 0.95 \\ 
\hline
$a''$ & 1.15 & 1.15 & 1.15 & 0.95 & 0.95 & 0.95 \\ 
\hline
$b'$  & 0.04 & 0.008 & 0.045 & 0.055 & 0.002 & 0.025 \\ 
\hline
$b''$ & -0.45 & 0.002 & 0.2 & -0.15 & 0.002 & 0.02 \\ 
\hline
\end{tabular}
\end{table*}

Our equivalent circuit approach produced analytical results close to those of the numerically calculated diode resistance, diode current, and metasurface transmittance. A conventional approach using a constant diode resistance value led to severe deviations, especially if the input power level was dynamically changed. The close agreement between our approach and the numerical results was obtained by using two mathematical solutions, namely, the Maclaurin expansion and Wright omega functions, which enabled us to find the approximated power- and time-dependent resistance of the diodes. This close agreement made accurate prediction of the diode current and the metasurface transmittance possible even under various conditions, including various input powers, different metasurfaces, and other frequencies. The proposed concept could be further extended to explain more complicated scenarios. For instance, Appendix B introduced the analytical $R_j$ and $X$ of capacitor-based waveform-selective metasurfaces under repeated pulses. Additionally, the duty cycle of repeated pulses could arbitrarily be changed by properly setting up the initial conditions of the current in $R_j$ and $X$ (or the initial conditions of the voltages in capacitor-based waveform-selective metasurfaces). Moreover, the same concept was applicable to more advanced types of waveform-selective metasurfaces, including those with a conducting geometry and circuit topology.\cite{wakatsuchi2015waveformSciRep, wakatsuchi2019waveform, takeshita2024frequency, wakatsuchi2015time}

\section{Conclusion}
We have presented an equivalent circuit approach to predict the power- and time-dependent electromagnetic responses of waveform-selective metasurfaces. Unlike a conventional approach using a constant value for a diode resistance, we introduced two mathematical solutions, specifically the Maclaurin series and the Wright omega function, which allowed us to derive the diode resistance, including the dependence on the power and duration of incident pulses. Therefore, the proposed method produced theoretically derived diode resistance, diode current, and metasurface transmittance, each of which closely agreed with the numerically derived results, whereas severe deviation appeared in the conventional method using a fixed diode resistance. We showed that the proposed concept could be extended to other scenarios, including those involving repeated pulses, different metasurfaces, and nonresonant frequencies. The presented approach enabled rapid optimization of waveform-selective metasurfaces without time-consuming numerical simulations, reducing the number of design cycles from more than hours to less than minutes. This capability is particularly valuable for emerging next-generation dynamic communication systems. Furthermore, the analytical nature of our solution would facilitate integration with system-level simulators and enable real-time adaptation algorithms for smart radio environments.

\begin{acknowledgments}

This work was supported in part by the Japan Science and Technology Agency (JST) under the Precursory Research for Embryonic Science and Technology (PRESTO) No.\ JPMJPR193A, under the Fusion Oriented Research for Disruptive Science and Technology (FOREST) No.\ JPMJFR222T, under the Adopting Sustainable Partnerships for Innovative Research Ecosystem (ASPIRE) No.\ JPMJAP2431, the Japan Society for the Promotion of Science (JSPS) under the Grant-in-Aid for Scientific Research (A) No.\ 23H00470, the Ministry of Internal Affairs and Communications (MIC) under the Fundamental Technologies for Sustainable Efficient Radio Wave Use R\&D Project (FORWARD) No.\ JPMI250610002, and the National Institute of Information and Communications Technology (NICT), Japan under the commissioned research No.\ JPJ012368C06201.
\end{acknowledgments}

\section*{Data Availability Statement}

The data that support the findings of this study are available from the corresponding author upon reasonable request.

\renewcommand{\thefigure}{S\arabic{figure}}
\setcounter{figure}{0}
\renewcommand{\thetable}{S\arabic{table}}
\setcounter{table}{0}

\section*{Appendix A: Derivation of Eq.\ (\ref{eq6:Laurent})}
Eq.\ (\ref{eq6:Laurent}) was obtained in the following manner. First, by using $X_0=(2R_j+2R_s+R_L+Z_0)t/L$, Eq.\ (\ref{eq6:Laurent}) could be rearranged to
\begin{eqnarray}
1-e^{X_0} \approx \frac{1}{\frac{1}{X_0}+1}=\frac{X_0}{1+X_0}.
\end{eqnarray}
Thus, we approximated a saturation curve related to an exponential function by another saturation curve related to a hyperbolic function. According to the Maclaurin series,
\begin{eqnarray}
e^{-X_0}=1-X_0+\frac{X_0^2}{2!}+\cdots .
\end{eqnarray}
Therefore, when $X_0$ was sufficiently small, $1-e^{X_0}$ approached 0 with a gradient of 0 at $X_0=0$. However, when $X_0 \rightarrow \infty$, $1-e^{X_0}$ approached 1. While satisfying these two conditions, we approximated $1-e^{X_0}$ by a rational function, namely, $(p_0+p_1X_0)/(q_0+q_1X_0)$, where $p_0$, $p_1$, $q_0$, and $q_1$ represent constants to be obtained. By considering the above values of $1-e^{X_0}$ at $X_0=0$ and $X_0\rightarrow \infty$, we obtained $p_0=0$ and $p_1=q_0=q_1$. Therefore, $X_0/(1+X_0)$ satisfied this condition and the approximation of Eq.\ (\ref{eq6:Laurent}).

\section*{Appendix B: $R_j$ and $X$ of capacitor-based waveform-selective metasurfaces under repeated pulse scenarios}
We derived Eqs.\ (\ref{eq10xx:ON}) to (\ref{eq11xx:OFF2}) for the $R_j$ and $X$ values of inductor-based waveform-selective metasurfaces under repeated pulse scenarios. By adopting a similar approach, $R_j$ and $X$ could be obtained for inductor-based waveform-selective metasurfaces under repeated pulse scenarios. Specifically, Eqs.\ (\ref{eq20:Rj}) and (\ref{eq21:WOF}) could be rearranged in on states (i.e., during a pulse) to
\begin{widetext}
\begin{eqnarray}
R_j=\frac{1}{2}\left( \frac{q(2(CR_C+t)V_{inc}-CR_CV')(tR_C+(2R_s+Z_0)(CR_C+t))}{2nKTX(CR_C+t)^2}-\frac{tR_C}{CR_C+t} -2R_s-Z_0\right),\label{eq20:Rjon}\\
X = \omega \left( \frac{q(2(CR_C+t)V_{inc}-CR_CV')}{2nKT(CR_C+t)}-\ln{\frac{2nKT(CR_C+t)}{qI_S\left( tR_C+(2R_s+Z_0)(CR_C+t)\right)}} \right),
\label{eq20:Xon}
\end{eqnarray}
\end{widetext}
where $V'$ represents the initial value of the capacitor voltage. With respect to the off states (i.e., between pulses), a closed-circuit loop was established between the parallel $C$ and $R_C$, implying that $R_j$ approached $\infty$. By alternately using this condition and Eqs.\ (\ref{eq20:Rjon}) and (\ref{eq20:Xon}), the behavior of capacitor-based waveform-selective metasurfaces could be estimated.

\begin{thebibliography}{50}%
\makeatletter
\providecommand \@ifxundefined [1]{%
 \@ifx{#1\undefined}
}%
\providecommand \@ifnum [1]{%
 \ifnum #1\expandafter \@firstoftwo
 \else \expandafter \@secondoftwo
 \fi
}%
\providecommand \@ifx [1]{%
 \ifx #1\expandafter \@firstoftwo
 \else \expandafter \@secondoftwo
 \fi
}%
\providecommand \natexlab [1]{#1}%
\providecommand \enquote  [1]{``#1''}%
\providecommand \bibnamefont  [1]{#1}%
\providecommand \bibfnamefont [1]{#1}%
\providecommand \citenamefont [1]{#1}%
\providecommand \href@noop [0]{\@secondoftwo}%
\providecommand \href [0]{\begingroup \@sanitize@url \@href}%
\providecommand \@href[1]{\@@startlink{#1}\@@href}%
\providecommand \@@href[1]{\endgroup#1\@@endlink}%
\providecommand \@sanitize@url [0]{\catcode `\\12\catcode `\$12\catcode `\&12\catcode `\#12\catcode `\^12\catcode `\_12\catcode `\%12\relax}%
\providecommand \@@startlink[1]{}%
\providecommand \@@endlink[0]{}%
\providecommand \url  [0]{\begingroup\@sanitize@url \@url }%
\providecommand \@url [1]{\endgroup\@href {#1}{\urlprefix }}%
\providecommand \urlprefix  [0]{URL }%
\providecommand \Eprint [0]{\href }%
\providecommand \doibase [0]{http://dx.doi.org/}%
\providecommand \selectlanguage [0]{\@gobble}%
\providecommand \bibinfo  [0]{\@secondoftwo}%
\providecommand \bibfield  [0]{\@secondoftwo}%
\providecommand \translation [1]{[#1]}%
\providecommand \BibitemOpen [0]{}%
\providecommand \bibitemStop [0]{}%
\providecommand \bibitemNoStop [0]{.\EOS\space}%
\providecommand \EOS [0]{\spacefactor3000\relax}%
\providecommand \BibitemShut  [1]{\csname bibitem#1\endcsname}%
\let\auto@bib@innerbib\@empty
\bibitem [{\citenamefont {Smith}\ \emph {et~al.}(2000)\citenamefont {Smith}, \citenamefont {Padilla}, \citenamefont {Vier}, \citenamefont {Nemat-Nasser},\ and\ \citenamefont {Schultz}}]{smithDNG1D}%
  \BibitemOpen
  \bibfield  {author} {\bibinfo {author} {\bibfnamefont {D.~R.}\ \bibnamefont {Smith}}, \bibinfo {author} {\bibfnamefont {W.~J.}\ \bibnamefont {Padilla}}, \bibinfo {author} {\bibfnamefont {D.~C.}\ \bibnamefont {Vier}}, \bibinfo {author} {\bibfnamefont {S.~C.}\ \bibnamefont {Nemat-Nasser}}, \ and\ \bibinfo {author} {\bibfnamefont {S.}~\bibnamefont {Schultz}},\ }\bibfield  {title} {\enquote {\bibinfo {title} {Composite medium with simultaneously negative permeability and permittivity},}\ }\href@noop {} {\bibfield  {journal} {\bibinfo  {journal} {Phys.\ Rev.\ Lett.}\ }\textbf {\bibinfo {volume} {84}},\ \bibinfo {pages} {4184--4187} (\bibinfo {year} {2000})}\BibitemShut {NoStop}%
\bibitem [{\citenamefont {Shelby}, \citenamefont {Smith},\ and\ \citenamefont {Schultz}(2001)}]{smithDNG2D2}%
  \BibitemOpen
  \bibfield  {author} {\bibinfo {author} {\bibfnamefont {R.~A.}\ \bibnamefont {Shelby}}, \bibinfo {author} {\bibfnamefont {D.~R.}\ \bibnamefont {Smith}}, \ and\ \bibinfo {author} {\bibfnamefont {S.}~\bibnamefont {Schultz}},\ }\bibfield  {title} {\enquote {\bibinfo {title} {Experimental verification of a negative index of refraction},}\ }\href@noop {} {\bibfield  {journal} {\bibinfo  {journal} {Science}\ }\textbf {\bibinfo {volume} {292}},\ \bibinfo {pages} {77--79} (\bibinfo {year} {2001})}\BibitemShut {NoStop}%
\bibitem [{\citenamefont {Sievenpiper}\ \emph {et~al.}(1999)\citenamefont {Sievenpiper}, \citenamefont {Zhang}, \citenamefont {Broas}, \citenamefont {Alex$\mathrm{\acute{o}}$polous},\ and\ \citenamefont {Yablonovitch}}]{EBGdevelopment}%
  \BibitemOpen
  \bibfield  {author} {\bibinfo {author} {\bibfnamefont {D.}~\bibnamefont {Sievenpiper}}, \bibinfo {author} {\bibfnamefont {L.}~\bibnamefont {Zhang}}, \bibinfo {author} {\bibfnamefont {R.~F.~J.}\ \bibnamefont {Broas}}, \bibinfo {author} {\bibfnamefont {N.~G.}\ \bibnamefont {Alex$\mathrm{\acute{o}}$polous}}, \ and\ \bibinfo {author} {\bibfnamefont {E.}~\bibnamefont {Yablonovitch}},\ }\bibfield  {title} {\enquote {\bibinfo {title} {High-impedance electromagnetic surfaces with a forbidden frequency band},}\ }\href@noop {} {\bibfield  {journal} {\bibinfo  {journal} {IEEE Trans. Microw. Theory Tech.}\ }\textbf {\bibinfo {volume} {47}},\ \bibinfo {pages} {2059--2074} (\bibinfo {year} {1999})}\BibitemShut {NoStop}%
\bibitem [{\citenamefont {Caloz}\ and\ \citenamefont {Itoh}(2006)}]{calozBook}%
  \BibitemOpen
  \bibfield  {author} {\bibinfo {author} {\bibfnamefont {C.}~\bibnamefont {Caloz}}\ and\ \bibinfo {author} {\bibfnamefont {T.}~\bibnamefont {Itoh}},\ }\href@noop {} {\emph {\bibinfo {title} {Electromagnetic metamaterials: transmission line theory and microwave applications}}}\ (\bibinfo  {publisher} {Wiley--IEEE Press},\ \bibinfo {address} {Hoboken, NJ},\ \bibinfo {year} {2006})\BibitemShut {NoStop}%
\bibitem [{\citenamefont {Engheta}\ and\ \citenamefont {Ziolkowski}(2006)}]{MTMbookEngheta}%
  \BibitemOpen
  \bibfield  {author} {\bibinfo {author} {\bibfnamefont {N.}~\bibnamefont {Engheta}}\ and\ \bibinfo {author} {\bibfnamefont {R.}~\bibnamefont {Ziolkowski}},\ }\href@noop {} {\emph {\bibinfo {title} {Metamaterials physics and engineering explorations}}}\ (\bibinfo  {publisher} {IEEE press, John Wiley \& Sons},\ \bibinfo {address} {Piscataway, NJ},\ \bibinfo {year} {2006})\BibitemShut {NoStop}%
\bibitem [{\citenamefont {Ziolkowski}(2004)}]{ziolkowski2004propagation}%
  \BibitemOpen
  \bibfield  {author} {\bibinfo {author} {\bibfnamefont {R.~W.}\ \bibnamefont {Ziolkowski}},\ }\bibfield  {title} {\enquote {\bibinfo {title} {Propagation in and scattering from a matched metamaterial having a zero index of refraction},}\ }\href@noop {} {\bibfield  {journal} {\bibinfo  {journal} {Phys.\ Rev.\ E}\ }\textbf {\bibinfo {volume} {70}},\ \bibinfo {pages} {046608} (\bibinfo {year} {2004})}\BibitemShut {NoStop}%
\bibitem [{\citenamefont {Liberal}\ and\ \citenamefont {Engheta}(2017)}]{liberal2017near}%
  \BibitemOpen
  \bibfield  {author} {\bibinfo {author} {\bibfnamefont {I.}~\bibnamefont {Liberal}}\ and\ \bibinfo {author} {\bibfnamefont {N.}~\bibnamefont {Engheta}},\ }\bibfield  {title} {\enquote {\bibinfo {title} {Near-zero refractive index photonics},}\ }\href@noop {} {\bibfield  {journal} {\bibinfo  {journal} {Nat.\ Photonics}\ }\textbf {\bibinfo {volume} {11}},\ \bibinfo {pages} {149--158} (\bibinfo {year} {2017})}\BibitemShut {NoStop}%
\bibitem [{\citenamefont {Yu}\ \emph {et~al.}(2011)\citenamefont {Yu}, \citenamefont {Genevet}, \citenamefont {Kats}, \citenamefont {Aieta}, \citenamefont {Tetienne}, \citenamefont {Capasso},\ and\ \citenamefont {Gaburro}}]{yu2011light}%
  \BibitemOpen
  \bibfield  {author} {\bibinfo {author} {\bibfnamefont {N.}~\bibnamefont {Yu}}, \bibinfo {author} {\bibfnamefont {P.}~\bibnamefont {Genevet}}, \bibinfo {author} {\bibfnamefont {M.~A.}\ \bibnamefont {Kats}}, \bibinfo {author} {\bibfnamefont {F.}~\bibnamefont {Aieta}}, \bibinfo {author} {\bibfnamefont {J.-P.}\ \bibnamefont {Tetienne}}, \bibinfo {author} {\bibfnamefont {F.}~\bibnamefont {Capasso}}, \ and\ \bibinfo {author} {\bibfnamefont {Z.}~\bibnamefont {Gaburro}},\ }\bibfield  {title} {\enquote {\bibinfo {title} {Light propagation with phase discontinuities: generalized laws of reflection and refraction},}\ }\href@noop {} {\bibfield  {journal} {\bibinfo  {journal} {Science}\ }\textbf {\bibinfo {volume} {334}},\ \bibinfo {pages} {333--337} (\bibinfo {year} {2011})}\BibitemShut {NoStop}%
\bibitem [{\citenamefont {Yu}\ and\ \citenamefont {Capasso}(2014)}]{yu2014flat}%
  \BibitemOpen
  \bibfield  {author} {\bibinfo {author} {\bibfnamefont {N.}~\bibnamefont {Yu}}\ and\ \bibinfo {author} {\bibfnamefont {F.}~\bibnamefont {Capasso}},\ }\bibfield  {title} {\enquote {\bibinfo {title} {Flat optics with designer metasurfaces},}\ }\href@noop {} {\bibfield  {journal} {\bibinfo  {journal} {Nat.\ Mater.}\ }\textbf {\bibinfo {volume} {13}},\ \bibinfo {pages} {139--150} (\bibinfo {year} {2014})}\BibitemShut {NoStop}%
\bibitem [{\citenamefont {Pendry}(2000)}]{pendryperfetLenses}%
  \BibitemOpen
  \bibfield  {author} {\bibinfo {author} {\bibfnamefont {J.~B.}\ \bibnamefont {Pendry}},\ }\bibfield  {title} {\enquote {\bibinfo {title} {Negative refraction makes a perfect lens},}\ }\href@noop {} {\bibfield  {journal} {\bibinfo  {journal} {Phys.\ Rev.\ Lett.}\ }\textbf {\bibinfo {volume} {85}},\ \bibinfo {pages} {3966--3969} (\bibinfo {year} {2000})}\BibitemShut {NoStop}%
\bibitem [{\citenamefont {Fang}\ \emph {et~al.}(2005)\citenamefont {Fang}, \citenamefont {Lee}, \citenamefont {Sun},\ and\ \citenamefont {Zhang}}]{fangSuperlens}%
  \BibitemOpen
  \bibfield  {author} {\bibinfo {author} {\bibfnamefont {N.}~\bibnamefont {Fang}}, \bibinfo {author} {\bibfnamefont {H.}~\bibnamefont {Lee}}, \bibinfo {author} {\bibfnamefont {C.}~\bibnamefont {Sun}}, \ and\ \bibinfo {author} {\bibfnamefont {X.}~\bibnamefont {Zhang}},\ }\bibfield  {title} {\enquote {\bibinfo {title} {Sub--diffraction--limited optical imaging with a silver superlens},}\ }\href@noop {} {\bibfield  {journal} {\bibinfo  {journal} {Science}\ }\textbf {\bibinfo {volume} {308}},\ \bibinfo {pages} {534--537} (\bibinfo {year} {2005})}\BibitemShut {NoStop}%
\bibitem [{\citenamefont {Grbic}, \citenamefont {Jiang},\ and\ \citenamefont {Merlin}(2008)}]{grbic2008near}%
  \BibitemOpen
  \bibfield  {author} {\bibinfo {author} {\bibfnamefont {A.}~\bibnamefont {Grbic}}, \bibinfo {author} {\bibfnamefont {L.}~\bibnamefont {Jiang}}, \ and\ \bibinfo {author} {\bibfnamefont {R.}~\bibnamefont {Merlin}},\ }\bibfield  {title} {\enquote {\bibinfo {title} {Near-field plates: subdiffraction focusing with patterned surfaces},}\ }\href@noop {} {\bibfield  {journal} {\bibinfo  {journal} {Science}\ }\textbf {\bibinfo {volume} {320}},\ \bibinfo {pages} {511--513} (\bibinfo {year} {2008})}\BibitemShut {NoStop}%
\bibitem [{\citenamefont {Pendry}, \citenamefont {Schurig},\ and\ \citenamefont {Smith}(2006)}]{pendryCloaking}%
  \BibitemOpen
  \bibfield  {author} {\bibinfo {author} {\bibfnamefont {J.~B.}\ \bibnamefont {Pendry}}, \bibinfo {author} {\bibfnamefont {D.}~\bibnamefont {Schurig}}, \ and\ \bibinfo {author} {\bibfnamefont {D.~R.}\ \bibnamefont {Smith}},\ }\bibfield  {title} {\enquote {\bibinfo {title} {Controlling electromagnetic fields},}\ }\href@noop {} {\bibfield  {journal} {\bibinfo  {journal} {Science}\ }\textbf {\bibinfo {volume} {312}},\ \bibinfo {pages} {1780--1782} (\bibinfo {year} {2006})}\BibitemShut {NoStop}%
\bibitem [{\citenamefont {Schurig}\ \emph {et~al.}(2006)\citenamefont {Schurig}, \citenamefont {Mock}, \citenamefont {Justice}, \citenamefont {Cummer}, \citenamefont {Pendry}, \citenamefont {Starr},\ and\ \citenamefont {Smith}}]{schurig2006metamaterial}%
  \BibitemOpen
  \bibfield  {author} {\bibinfo {author} {\bibfnamefont {D.}~\bibnamefont {Schurig}}, \bibinfo {author} {\bibfnamefont {J.}~\bibnamefont {Mock}}, \bibinfo {author} {\bibfnamefont {B.}~\bibnamefont {Justice}}, \bibinfo {author} {\bibfnamefont {S.~A.}\ \bibnamefont {Cummer}}, \bibinfo {author} {\bibfnamefont {J.~B.}\ \bibnamefont {Pendry}}, \bibinfo {author} {\bibfnamefont {A.}~\bibnamefont {Starr}}, \ and\ \bibinfo {author} {\bibfnamefont {D.}~\bibnamefont {Smith}},\ }\bibfield  {title} {\enquote {\bibinfo {title} {Metamaterial electromagnetic cloak at microwave frequencies},}\ }\href@noop {} {\bibfield  {journal} {\bibinfo  {journal} {Science}\ }\textbf {\bibinfo {volume} {314}},\ \bibinfo {pages} {977--980} (\bibinfo {year} {2006})}\BibitemShut {NoStop}%
\bibitem [{\citenamefont {Al$\mathrm{\grave{u}}$}\ and\ \citenamefont {Engheta}(2005)}]{enghetaCloaking}%
  \BibitemOpen
  \bibfield  {author} {\bibinfo {author} {\bibfnamefont {A.}~\bibnamefont {Al$\mathrm{\grave{u}}$}}\ and\ \bibinfo {author} {\bibfnamefont {N.}~\bibnamefont {Engheta}},\ }\bibfield  {title} {\enquote {\bibinfo {title} {Achieving transparency with plasmonic and metamaterial coatings},}\ }\href@noop {} {\bibfield  {journal} {\bibinfo  {journal} {Phys.\ Rev.\ E}\ }\textbf {\bibinfo {volume} {72}},\ \bibinfo {pages} {016623} (\bibinfo {year} {2005})}\BibitemShut {NoStop}%
\bibitem [{\citenamefont {Landy}\ \emph {et~al.}(2008)\citenamefont {Landy}, \citenamefont {Sajuyigbe}, \citenamefont {Mock}, \citenamefont {Smith},\ and\ \citenamefont {Padilla}}]{mtmAbsPRLpadilla}%
  \BibitemOpen
  \bibfield  {author} {\bibinfo {author} {\bibfnamefont {N.~I.}\ \bibnamefont {Landy}}, \bibinfo {author} {\bibfnamefont {S.}~\bibnamefont {Sajuyigbe}}, \bibinfo {author} {\bibfnamefont {J.~J.}\ \bibnamefont {Mock}}, \bibinfo {author} {\bibfnamefont {D.~R.}\ \bibnamefont {Smith}}, \ and\ \bibinfo {author} {\bibfnamefont {W.~J.}\ \bibnamefont {Padilla}},\ }\bibfield  {title} {\enquote {\bibinfo {title} {Perfect metamaterial absorber},}\ }\href@noop {} {\bibfield  {journal} {\bibinfo  {journal} {Phys.\ Rev.\ Lett.}\ }\textbf {\bibinfo {volume} {100}},\ \bibinfo {pages} {207402} (\bibinfo {year} {2008})}\BibitemShut {NoStop}%
\bibitem [{\citenamefont {Wakatsuchi}\ \emph {et~al.}(2010)\citenamefont {Wakatsuchi}, \citenamefont {Greedy}, \citenamefont {Christopoulos},\ and\ \citenamefont {Paul}}]{My1stAbsPaper}%
  \BibitemOpen
  \bibfield  {author} {\bibinfo {author} {\bibfnamefont {H.}~\bibnamefont {Wakatsuchi}}, \bibinfo {author} {\bibfnamefont {S.}~\bibnamefont {Greedy}}, \bibinfo {author} {\bibfnamefont {C.}~\bibnamefont {Christopoulos}}, \ and\ \bibinfo {author} {\bibfnamefont {J.}~\bibnamefont {Paul}},\ }\bibfield  {title} {\enquote {\bibinfo {title} {Customised broadband metamaterial absorbers for arbitrary polarisation},}\ }\href@noop {} {\bibfield  {journal} {\bibinfo  {journal} {Opt.\ Express}\ }\textbf {\bibinfo {volume} {18}},\ \bibinfo {pages} {22187--22198} (\bibinfo {year} {2010})}\BibitemShut {NoStop}%
\bibitem [{\citenamefont {Silva}\ \emph {et~al.}(2014)\citenamefont {Silva}, \citenamefont {Monticone}, \citenamefont {Castaldi}, \citenamefont {Galdi}, \citenamefont {Al{\`u}},\ and\ \citenamefont {Engheta}}]{silva2014performing}%
  \BibitemOpen
  \bibfield  {author} {\bibinfo {author} {\bibfnamefont {A.}~\bibnamefont {Silva}}, \bibinfo {author} {\bibfnamefont {F.}~\bibnamefont {Monticone}}, \bibinfo {author} {\bibfnamefont {G.}~\bibnamefont {Castaldi}}, \bibinfo {author} {\bibfnamefont {V.}~\bibnamefont {Galdi}}, \bibinfo {author} {\bibfnamefont {A.}~\bibnamefont {Al{\`u}}}, \ and\ \bibinfo {author} {\bibfnamefont {N.}~\bibnamefont {Engheta}},\ }\bibfield  {title} {\enquote {\bibinfo {title} {Performing mathematical operations with metamaterials},}\ }\href@noop {} {\bibfield  {journal} {\bibinfo  {journal} {Science}\ }\textbf {\bibinfo {volume} {343}},\ \bibinfo {pages} {160--163} (\bibinfo {year} {2014})}\BibitemShut {NoStop}%
\bibitem [{\citenamefont {Mohammadi~Estakhri}, \citenamefont {Edwards},\ and\ \citenamefont {Engheta}(2019)}]{mohammadi2019inverse}%
  \BibitemOpen
  \bibfield  {author} {\bibinfo {author} {\bibfnamefont {N.}~\bibnamefont {Mohammadi~Estakhri}}, \bibinfo {author} {\bibfnamefont {B.}~\bibnamefont {Edwards}}, \ and\ \bibinfo {author} {\bibfnamefont {N.}~\bibnamefont {Engheta}},\ }\bibfield  {title} {\enquote {\bibinfo {title} {Inverse-designed metastructures that solve equations},}\ }\href@noop {} {\bibfield  {journal} {\bibinfo  {journal} {Science}\ }\textbf {\bibinfo {volume} {363}},\ \bibinfo {pages} {1333--1338} (\bibinfo {year} {2019})}\BibitemShut {NoStop}%
\bibitem [{\citenamefont {Zangeneh-Nejad}\ \emph {et~al.}(2021)\citenamefont {Zangeneh-Nejad}, \citenamefont {Sounas}, \citenamefont {Al{\`u}},\ and\ \citenamefont {Fleury}}]{zangeneh2021analogue}%
  \BibitemOpen
  \bibfield  {author} {\bibinfo {author} {\bibfnamefont {F.}~\bibnamefont {Zangeneh-Nejad}}, \bibinfo {author} {\bibfnamefont {D.~L.}\ \bibnamefont {Sounas}}, \bibinfo {author} {\bibfnamefont {A.}~\bibnamefont {Al{\`u}}}, \ and\ \bibinfo {author} {\bibfnamefont {R.}~\bibnamefont {Fleury}},\ }\bibfield  {title} {\enquote {\bibinfo {title} {Analogue computing with metamaterials},}\ }\href@noop {} {\bibfield  {journal} {\bibinfo  {journal} {Nat.\ Rev.\ Mater.}\ }\textbf {\bibinfo {volume} {6}},\ \bibinfo {pages} {207--225} (\bibinfo {year} {2021})}\BibitemShut {NoStop}%
\bibitem [{\citenamefont {Ziolkowski}\ and\ \citenamefont {Erentok}(2006)}]{ziolkowski2006metamaterial}%
  \BibitemOpen
  \bibfield  {author} {\bibinfo {author} {\bibfnamefont {R.~W.}\ \bibnamefont {Ziolkowski}}\ and\ \bibinfo {author} {\bibfnamefont {A.}~\bibnamefont {Erentok}},\ }\bibfield  {title} {\enquote {\bibinfo {title} {Metamaterial-based efficient electrically small antennas},}\ }\href@noop {} {\bibfield  {journal} {\bibinfo  {journal} {IEEE Trans.\ Antennas Propag.}\ }\textbf {\bibinfo {volume} {54}},\ \bibinfo {pages} {2113--2130} (\bibinfo {year} {2006})}\BibitemShut {NoStop}%
\bibitem [{\citenamefont {Ziolkowski}, \citenamefont {Jin},\ and\ \citenamefont {Lin}(2011)}]{ziolkowski2011metamaterial}%
  \BibitemOpen
  \bibfield  {author} {\bibinfo {author} {\bibfnamefont {R.~W.}\ \bibnamefont {Ziolkowski}}, \bibinfo {author} {\bibfnamefont {P.}~\bibnamefont {Jin}}, \ and\ \bibinfo {author} {\bibfnamefont {C.-C.}\ \bibnamefont {Lin}},\ }\bibfield  {title} {\enquote {\bibinfo {title} {Metamaterial-inspired engineering of antennas},}\ }\href@noop {} {\bibfield  {journal} {\bibinfo  {journal} {Proc.\ IEEE}\ }\textbf {\bibinfo {volume} {99}},\ \bibinfo {pages} {1720--1731} (\bibinfo {year} {2011})}\BibitemShut {NoStop}%
\bibitem [{\citenamefont {Wu}\ and\ \citenamefont {Zhang}(2019)}]{wu2019towards}%
  \BibitemOpen
  \bibfield  {author} {\bibinfo {author} {\bibfnamefont {Q.}~\bibnamefont {Wu}}\ and\ \bibinfo {author} {\bibfnamefont {R.}~\bibnamefont {Zhang}},\ }\bibfield  {title} {\enquote {\bibinfo {title} {Towards smart and reconfigurable environment: intelligent reflecting surface aided wireless network},}\ }\href@noop {} {\bibfield  {journal} {\bibinfo  {journal} {IEEE Commun.\ Mag.}\ }\textbf {\bibinfo {volume} {58}},\ \bibinfo {pages} {106--112} (\bibinfo {year} {2019})}\BibitemShut {NoStop}%
\bibitem [{\citenamefont {Sugiura}\ \emph {et~al.}(2021)\citenamefont {Sugiura}, \citenamefont {Kawai}, \citenamefont {Matsui}, \citenamefont {Lee},\ and\ \citenamefont {Iizuka}}]{sugiura2021joint}%
  \BibitemOpen
  \bibfield  {author} {\bibinfo {author} {\bibfnamefont {S.}~\bibnamefont {Sugiura}}, \bibinfo {author} {\bibfnamefont {Y.}~\bibnamefont {Kawai}}, \bibinfo {author} {\bibfnamefont {T.}~\bibnamefont {Matsui}}, \bibinfo {author} {\bibfnamefont {T.}~\bibnamefont {Lee}}, \ and\ \bibinfo {author} {\bibfnamefont {H.}~\bibnamefont {Iizuka}},\ }\bibfield  {title} {\enquote {\bibinfo {title} {Joint beam and polarization forming of intelligent reflecting surfaces for wireless communications},}\ }\href@noop {} {\bibfield  {journal} {\bibinfo  {journal} {IEEE Trans.\ Veh.\ Technol.}\ }\textbf {\bibinfo {volume} {70}},\ \bibinfo {pages} {1648--1657} (\bibinfo {year} {2021})}\BibitemShut {NoStop}%
\bibitem [{\citenamefont {Ino}, \citenamefont {Wakatsuchi},\ and\ \citenamefont {Sugiura}(2023)}]{ino2023noncoherent}%
  \BibitemOpen
  \bibfield  {author} {\bibinfo {author} {\bibfnamefont {J.}~\bibnamefont {Ino}}, \bibinfo {author} {\bibfnamefont {H.}~\bibnamefont {Wakatsuchi}}, \ and\ \bibinfo {author} {\bibfnamefont {S.}~\bibnamefont {Sugiura}},\ }\bibfield  {title} {\enquote {\bibinfo {title} {Noncoherent reconfigurable intelligent surface with differential modulation and reflection pattern training},}\ }\href@noop {} {\bibfield  {journal} {\bibinfo  {journal} {IEEE Wirel.\ Commun.\ Lett.}\ }\textbf {\bibinfo {volume} {13}},\ \bibinfo {pages} {656--660} (\bibinfo {year} {2023})}\BibitemShut {NoStop}%
\bibitem [{\citenamefont {Takimoto}\ \emph {et~al.}(2025)\citenamefont {Takimoto}, \citenamefont {Njogu}, \citenamefont {Ramachandran}, \citenamefont {Kondo}, \citenamefont {Mori}, \citenamefont {Anzai},\ and\ \citenamefont {Wakatsuchi}}]{takimoto2025milli}%
  \BibitemOpen
  \bibfield  {author} {\bibinfo {author} {\bibfnamefont {K.}~\bibnamefont {Takimoto}}, \bibinfo {author} {\bibfnamefont {P.}~\bibnamefont {Njogu}}, \bibinfo {author} {\bibfnamefont {T.}~\bibnamefont {Ramachandran}}, \bibinfo {author} {\bibfnamefont {T.}~\bibnamefont {Kondo}}, \bibinfo {author} {\bibfnamefont {M.}~\bibnamefont {Mori}}, \bibinfo {author} {\bibfnamefont {D.}~\bibnamefont {Anzai}}, \ and\ \bibinfo {author} {\bibfnamefont {H.}~\bibnamefont {Wakatsuchi}},\ }\bibfield  {title} {\enquote {\bibinfo {title} {Millimeter-wave and sub-terahertz-band beamforming facilitated by paper-based passive intelligent reflecting surfaces},}\ }\href@noop {} {\bibfield  {journal} {\bibinfo  {journal} {Adv.\ Eng.\ Mater.}\ \textbf {\bibinfo {volume} {27}},\ \bibinfo {pages} {2500911}} (\bibinfo {year} {2025})}\BibitemShut {NoStop}%
\bibitem [{\citenamefont {Zhang}\ \emph {et~al.}(2018)\citenamefont {Zhang}, \citenamefont {Chen}, \citenamefont {Liu}, \citenamefont {Zhang}, \citenamefont {Zhao}, \citenamefont {Dai}, \citenamefont {Bai}, \citenamefont {Wan}, \citenamefont {Cheng}, \citenamefont {Castaldi} \emph {et~al.}}]{zhang2018space}%
  \BibitemOpen
  \bibfield  {author} {\bibinfo {author} {\bibfnamefont {L.}~\bibnamefont {Zhang}}, \bibinfo {author} {\bibfnamefont {X.~Q.}\ \bibnamefont {Chen}}, \bibinfo {author} {\bibfnamefont {S.}~\bibnamefont {Liu}}, \bibinfo {author} {\bibfnamefont {Q.}~\bibnamefont {Zhang}}, \bibinfo {author} {\bibfnamefont {J.}~\bibnamefont {Zhao}}, \bibinfo {author} {\bibfnamefont {J.~Y.}\ \bibnamefont {Dai}}, \bibinfo {author} {\bibfnamefont {G.~D.}\ \bibnamefont {Bai}}, \bibinfo {author} {\bibfnamefont {X.}~\bibnamefont {Wan}}, \bibinfo {author} {\bibfnamefont {Q.}~\bibnamefont {Cheng}}, \bibinfo {author} {\bibfnamefont {G.}~\bibnamefont {Castaldi}},  \emph {et~al.},\ }\bibfield  {title} {\enquote {\bibinfo {title} {Space-time-coding digital metasurfaces},}\ }\href@noop {} {\bibfield  {journal} {\bibinfo  {journal} {Nat.\ Commun.}\ }\textbf {\bibinfo {volume} {9}},\ \bibinfo {pages} {1--11} (\bibinfo {year} {2018})}\BibitemShut {NoStop}%
\bibitem [{\citenamefont {Di~Renzo}\ \emph {et~al.}(2020)\citenamefont {Di~Renzo}, \citenamefont {Zappone}, \citenamefont {Debbah}, \citenamefont {Alouini}, \citenamefont {Yuen}, \citenamefont {De~Rosny},\ and\ \citenamefont {Tretyakov}}]{di2020smart}%
  \BibitemOpen
  \bibfield  {author} {\bibinfo {author} {\bibfnamefont {M.}~\bibnamefont {Di~Renzo}}, \bibinfo {author} {\bibfnamefont {A.}~\bibnamefont {Zappone}}, \bibinfo {author} {\bibfnamefont {M.}~\bibnamefont {Debbah}}, \bibinfo {author} {\bibfnamefont {M.-S.}\ \bibnamefont {Alouini}}, \bibinfo {author} {\bibfnamefont {C.}~\bibnamefont {Yuen}}, \bibinfo {author} {\bibfnamefont {J.}~\bibnamefont {De~Rosny}}, \ and\ \bibinfo {author} {\bibfnamefont {S.}~\bibnamefont {Tretyakov}},\ }\bibfield  {title} {\enquote {\bibinfo {title} {Smart radio environments empowered by reconfigurable intelligent surfaces: how it works, state of research, and the road ahead},}\ }\href@noop {} {\bibfield  {journal} {\bibinfo  {journal} {IEEE J.\ Sel.\ Areas Commun.}\ }\textbf {\bibinfo {volume} {38}},\ \bibinfo {pages} {2450--2525} (\bibinfo {year} {2020})}\BibitemShut {NoStop}%
\bibitem [{\citenamefont {Dai}\ \emph {et~al.}(2020)\citenamefont {Dai}, \citenamefont {Wang}, \citenamefont {Wang}, \citenamefont {Yang}, \citenamefont {Tan}, \citenamefont {Bi}, \citenamefont {Xu}, \citenamefont {Yang}, \citenamefont {Chen}, \citenamefont {Di~Renzo} \emph {et~al.}}]{dai2020reconfigurable}%
  \BibitemOpen
  \bibfield  {author} {\bibinfo {author} {\bibfnamefont {L.}~\bibnamefont {Dai}}, \bibinfo {author} {\bibfnamefont {B.}~\bibnamefont {Wang}}, \bibinfo {author} {\bibfnamefont {M.}~\bibnamefont {Wang}}, \bibinfo {author} {\bibfnamefont {X.}~\bibnamefont {Yang}}, \bibinfo {author} {\bibfnamefont {J.}~\bibnamefont {Tan}}, \bibinfo {author} {\bibfnamefont {S.}~\bibnamefont {Bi}}, \bibinfo {author} {\bibfnamefont {S.}~\bibnamefont {Xu}}, \bibinfo {author} {\bibfnamefont {F.}~\bibnamefont {Yang}}, \bibinfo {author} {\bibfnamefont {Z.}~\bibnamefont {Chen}}, \bibinfo {author} {\bibfnamefont {M.}~\bibnamefont {Di~Renzo}},  \emph {et~al.},\ }\bibfield  {title} {\enquote {\bibinfo {title} {Reconfigurable intelligent surface-based wireless communications: antenna design, prototyping, and experimental results},}\ }\href@noop {} {\bibfield  {journal} {\bibinfo  {journal} {IEEE Access}\ }\textbf {\bibinfo {volume} {8}},\ \bibinfo {pages} {45913--45923} (\bibinfo {year} {2020})}\BibitemShut {NoStop}%
\bibitem [{\citenamefont {Gradoni}\ \emph {et~al.}(2022)\citenamefont {Gradoni}, \citenamefont {Di~Renzo}, \citenamefont {Diaz-Rubio}, \citenamefont {Tretyakov}, \citenamefont {Caloz}, \citenamefont {Peng}, \citenamefont {Alu}, \citenamefont {Lerosey}, \citenamefont {Fink}, \citenamefont {Galdi} \emph {et~al.}}]{gradoni1smart}%
  \BibitemOpen
  \bibfield  {author} {\bibinfo {author} {\bibfnamefont {G.}~\bibnamefont {Gradoni}}, \bibinfo {author} {\bibfnamefont {M.}~\bibnamefont {Di~Renzo}}, \bibinfo {author} {\bibfnamefont {A.}~\bibnamefont {Diaz-Rubio}}, \bibinfo {author} {\bibfnamefont {S.}~\bibnamefont {Tretyakov}}, \bibinfo {author} {\bibfnamefont {C.}~\bibnamefont {Caloz}}, \bibinfo {author} {\bibfnamefont {Z.}~\bibnamefont {Peng}}, \bibinfo {author} {\bibfnamefont {A.}~\bibnamefont {Alu}}, \bibinfo {author} {\bibfnamefont {G.}~\bibnamefont {Lerosey}}, \bibinfo {author} {\bibfnamefont {M.}~\bibnamefont {Fink}}, \bibinfo {author} {\bibfnamefont {V.}~\bibnamefont {Galdi}},  \emph {et~al.},\ }\bibfield  {title} {\enquote {\bibinfo {title} {Smart surface radio environments},}\ }\href@noop {} {\bibfield  {journal} {\bibinfo  {journal} {Reviews of Electromagnetics}\ }\textbf {\bibinfo {volume} {1}} (\bibinfo {year} {2022})}\BibitemShut {NoStop}%
\bibitem [{\citenamefont {Wakatsuchi}\ \emph {et~al.}(2013)\citenamefont {Wakatsuchi}, \citenamefont {Kim}, \citenamefont {Rushton},\ and\ \citenamefont {Sievenpiper}}]{wakatsuchi2013waveform}%
  \BibitemOpen
  \bibfield  {author} {\bibinfo {author} {\bibfnamefont {H.}~\bibnamefont {Wakatsuchi}}, \bibinfo {author} {\bibfnamefont {S.}~\bibnamefont {Kim}}, \bibinfo {author} {\bibfnamefont {J.~J.}\ \bibnamefont {Rushton}}, \ and\ \bibinfo {author} {\bibfnamefont {D.~F.}\ \bibnamefont {Sievenpiper}},\ }\bibfield  {title} {\enquote {\bibinfo {title} {Waveform-dependent absorbing metasurfaces},}\ }\href@noop {} {\bibfield  {journal} {\bibinfo  {journal} {Phys.\ Rev.\ Lett.}\ }\textbf {\bibinfo {volume} {111}},\ \bibinfo {pages} {245501} (\bibinfo {year} {2013})}\BibitemShut {NoStop}%
\bibitem [{\citenamefont {Eleftheriades}(2014)}]{eleftheriades2014electronics}%
  \BibitemOpen
  \bibfield  {author} {\bibinfo {author} {\bibfnamefont {G.~V.}\ \bibnamefont {Eleftheriades}},\ }\bibfield  {title} {\enquote {\bibinfo {title} {Electronics: protecting the weak from the strong},}\ }\href@noop {} {\bibfield  {journal} {\bibinfo  {journal} {Nature}\ }\textbf {\bibinfo {volume} {505}},\ \bibinfo {pages} {490--491} (\bibinfo {year} {2014})}\BibitemShut {NoStop}%
\bibitem [{\citenamefont {Wakatsuchi}\ \emph {et~al.}(2015)\citenamefont {Wakatsuchi}, \citenamefont {Anzai}, \citenamefont {Rushton}, \citenamefont {Gao}, \citenamefont {Kim},\ and\ \citenamefont {Sievenpiper}}]{wakatsuchi2015waveformSciRep}%
  \BibitemOpen
  \bibfield  {author} {\bibinfo {author} {\bibfnamefont {H.}~\bibnamefont {Wakatsuchi}}, \bibinfo {author} {\bibfnamefont {D.}~\bibnamefont {Anzai}}, \bibinfo {author} {\bibfnamefont {J.~J.}\ \bibnamefont {Rushton}}, \bibinfo {author} {\bibfnamefont {F.}~\bibnamefont {Gao}}, \bibinfo {author} {\bibfnamefont {S.}~\bibnamefont {Kim}}, \ and\ \bibinfo {author} {\bibfnamefont {D.~F.}\ \bibnamefont {Sievenpiper}},\ }\bibfield  {title} {\enquote {\bibinfo {title} {Waveform selectivity at the same frequency},}\ }\href@noop {} {\bibfield  {journal} {\bibinfo  {journal} {Sci.\ Rep.}\ }\textbf {\bibinfo {volume} {5}},\ \bibinfo {pages} {9639} (\bibinfo {year} {2015})}\BibitemShut {NoStop}%
\bibitem [{\citenamefont {Vellucci}\ \emph {et~al.}(2019)\citenamefont {Vellucci}, \citenamefont {Monti}, \citenamefont {Barbuto}, \citenamefont {Toscano},\ and\ \citenamefont {Bilotti}}]{vellucci2019waveform}%
  \BibitemOpen
  \bibfield  {author} {\bibinfo {author} {\bibfnamefont {S.}~\bibnamefont {Vellucci}}, \bibinfo {author} {\bibfnamefont {A.}~\bibnamefont {Monti}}, \bibinfo {author} {\bibfnamefont {M.}~\bibnamefont {Barbuto}}, \bibinfo {author} {\bibfnamefont {A.}~\bibnamefont {Toscano}}, \ and\ \bibinfo {author} {\bibfnamefont {F.}~\bibnamefont {Bilotti}},\ }\bibfield  {title} {\enquote {\bibinfo {title} {Waveform-selective mantle cloaks for intelligent antennas},}\ }\href@noop {} {\bibfield  {journal} {\bibinfo  {journal} {IEEE Trans. Antennas Propag.}\ }\textbf {\bibinfo {volume} {68}},\ \bibinfo {pages} {1717--1725} (\bibinfo {year} {2019})}\BibitemShut {NoStop}%
\bibitem [{\citenamefont {F.~Imani}\ and\ \citenamefont {Smith}(2020)}]{f2020temporal}%
  \BibitemOpen
  \bibfield  {author} {\bibinfo {author} {\bibfnamefont {M.}~\bibnamefont {F.~Imani}}\ and\ \bibinfo {author} {\bibfnamefont {D.~R.}\ \bibnamefont {Smith}},\ }\bibfield  {title} {\enquote {\bibinfo {title} {Temporal microwave ghost imaging using a reconfigurable disordered cavity},}\ }\href@noop {} {\bibfield  {journal} {\bibinfo  {journal} {Appl.\ Phys.\ Lett.}\ }\textbf {\bibinfo {volume} {116}},\ \bibinfo {pages} {054102} (\bibinfo {year} {2020})}\BibitemShut {NoStop}%
\bibitem [{\citenamefont {Takeshita}\ \emph {et~al.}(2024)\citenamefont {Takeshita}, \citenamefont {Fathnan}, \citenamefont {Nita}, \citenamefont {Nagata}, \citenamefont {Sugiura},\ and\ \citenamefont {Wakatsuchi}}]{takeshita2024frequency}%
  \BibitemOpen
  \bibfield  {author} {\bibinfo {author} {\bibfnamefont {H.}~\bibnamefont {Takeshita}}, \bibinfo {author} {\bibfnamefont {A.~A.}\ \bibnamefont {Fathnan}}, \bibinfo {author} {\bibfnamefont {D.}~\bibnamefont {Nita}}, \bibinfo {author} {\bibfnamefont {A.}~\bibnamefont {Nagata}}, \bibinfo {author} {\bibfnamefont {S.}~\bibnamefont {Sugiura}}, \ and\ \bibinfo {author} {\bibfnamefont {H.}~\bibnamefont {Wakatsuchi}},\ }\bibfield  {title} {\enquote {\bibinfo {title} {Frequency-hopping wave engineering with metasurfaces},}\ }\href@noop {} {\bibfield  {journal} {\bibinfo  {journal} {Nat.\ Commun.}\ }\textbf {\bibinfo {volume} {15}},\ \bibinfo {pages} {196} (\bibinfo {year} {2024})}\BibitemShut {NoStop}%
\bibitem [{\citenamefont {Wakatsuchi}, \citenamefont {Long},\ and\ \citenamefont {Sievenpiper}(2019)}]{wakatsuchi2019waveform}%
  \BibitemOpen
  \bibfield  {author} {\bibinfo {author} {\bibfnamefont {H.}~\bibnamefont {Wakatsuchi}}, \bibinfo {author} {\bibfnamefont {J.}~\bibnamefont {Long}}, \ and\ \bibinfo {author} {\bibfnamefont {D.~F.}\ \bibnamefont {Sievenpiper}},\ }\bibfield  {title} {\enquote {\bibinfo {title} {Waveform selective surfaces},}\ }\href@noop {} {\bibfield  {journal} {\bibinfo  {journal} {Adv.\ Funct.\ Mater.}\ }\textbf {\bibinfo {volume} {29}},\ \bibinfo {pages} {1806386} (\bibinfo {year} {2019})}\BibitemShut {NoStop}%
\bibitem [{\citenamefont {Ushikoshi}\ \emph {et~al.}(2023)\citenamefont {Ushikoshi}, \citenamefont {Higashiura}, \citenamefont {Tachi}, \citenamefont {Fathnan}, \citenamefont {Mahmood}, \citenamefont {Takeshita}, \citenamefont {Homma}, \citenamefont {Akram}, \citenamefont {Vellucci}, \citenamefont {Lee} \emph {et~al.}}]{ushikoshi2023pulse}%
  \BibitemOpen
  \bibfield  {author} {\bibinfo {author} {\bibfnamefont {D.}~\bibnamefont {Ushikoshi}}, \bibinfo {author} {\bibfnamefont {R.}~\bibnamefont {Higashiura}}, \bibinfo {author} {\bibfnamefont {K.}~\bibnamefont {Tachi}}, \bibinfo {author} {\bibfnamefont {A.~A.}\ \bibnamefont {Fathnan}}, \bibinfo {author} {\bibfnamefont {S.}~\bibnamefont {Mahmood}}, \bibinfo {author} {\bibfnamefont {H.}~\bibnamefont {Takeshita}}, \bibinfo {author} {\bibfnamefont {H.}~\bibnamefont {Homma}}, \bibinfo {author} {\bibfnamefont {M.~R.}\ \bibnamefont {Akram}}, \bibinfo {author} {\bibfnamefont {S.}~\bibnamefont {Vellucci}}, \bibinfo {author} {\bibfnamefont {J.}~\bibnamefont {Lee}},  \emph {et~al.},\ }\bibfield  {title} {\enquote {\bibinfo {title} {Pulse-driven self-reconfigurable meta-antennas},}\ }\href@noop {} {\bibfield  {journal} {\bibinfo  {journal} {Nat.\ Commun.}\ }\textbf {\bibinfo {volume} {14}},\ \bibinfo {pages} {633} (\bibinfo {year} {2023})}\BibitemShut {NoStop}%
\bibitem [{\citenamefont {Fathnan}\ \emph {et~al.}(2023)\citenamefont {Fathnan}, \citenamefont {Takimoto}, \citenamefont {Tanikawa}, \citenamefont {Nakamura}, \citenamefont {Sugiura},\ and\ \citenamefont {Wakatsuchi}}]{fathnan2023unsynchronized}%
  \BibitemOpen
  \bibfield  {author} {\bibinfo {author} {\bibfnamefont {A.~A.}\ \bibnamefont {Fathnan}}, \bibinfo {author} {\bibfnamefont {K.}~\bibnamefont {Takimoto}}, \bibinfo {author} {\bibfnamefont {M.}~\bibnamefont {Tanikawa}}, \bibinfo {author} {\bibfnamefont {K.}~\bibnamefont {Nakamura}}, \bibinfo {author} {\bibfnamefont {S.}~\bibnamefont {Sugiura}}, \ and\ \bibinfo {author} {\bibfnamefont {H.}~\bibnamefont {Wakatsuchi}},\ }\bibfield  {title} {\enquote {\bibinfo {title} {Unsynchronized reconfigurable intelligent surfaces with pulse-width-based design},}\ }\href@noop {} {\bibfield  {journal} {\bibinfo  {journal} {IEEE Trans.\ Veh.\ Technol.}\ }\textbf {\bibinfo {volume} {72}},\ \bibinfo {pages} {15103--15108} (\bibinfo {year} {2023})}\BibitemShut {NoStop}%
\bibitem [{\citenamefont {Baena}\ \emph {et~al.}(2005)\citenamefont {Baena}, \citenamefont {Bonache}, \citenamefont {Mart{\'\i}n}, \citenamefont {Sillero}, \citenamefont {Falcone}, \citenamefont {Lopetegi}, \citenamefont {Laso}, \citenamefont {Garcia-Garcia}, \citenamefont {Gil}, \citenamefont {Portillo} \emph {et~al.}}]{baena2005equivalent}%
  \BibitemOpen
  \bibfield  {author} {\bibinfo {author} {\bibfnamefont {J.~D.}\ \bibnamefont {Baena}}, \bibinfo {author} {\bibfnamefont {J.}~\bibnamefont {Bonache}}, \bibinfo {author} {\bibfnamefont {F.}~\bibnamefont {Mart{\'\i}n}}, \bibinfo {author} {\bibfnamefont {R.~M.}\ \bibnamefont {Sillero}}, \bibinfo {author} {\bibfnamefont {F.}~\bibnamefont {Falcone}}, \bibinfo {author} {\bibfnamefont {T.}~\bibnamefont {Lopetegi}}, \bibinfo {author} {\bibfnamefont {M.~A.}\ \bibnamefont {Laso}}, \bibinfo {author} {\bibfnamefont {J.}~\bibnamefont {Garcia-Garcia}}, \bibinfo {author} {\bibfnamefont {I.}~\bibnamefont {Gil}}, \bibinfo {author} {\bibfnamefont {M.~F.}\ \bibnamefont {Portillo}},  \emph {et~al.},\ }\bibfield  {title} {\enquote {\bibinfo {title} {Equivalent-circuit models for split-ring resonators and complementary split-ring resonators coupled to planar transmission lines},}\ }\href@noop {} {\bibfield  {journal} {\bibinfo  {journal} {IEEE Trans.\ Microw.\ Theory Tech.}\ }\textbf {\bibinfo {volume} {53}},\
  \bibinfo {pages} {1451--1461} (\bibinfo {year} {2005})}\BibitemShut {NoStop}%
\bibitem [{\citenamefont {Zhou}\ \emph {et~al.}(2006)\citenamefont {Zhou}, \citenamefont {Economou}, \citenamefont {Koschny},\ and\ \citenamefont {Soukoulis}}]{ZhouCWeq}%
  \BibitemOpen
  \bibfield  {author} {\bibinfo {author} {\bibfnamefont {J.}~\bibnamefont {Zhou}}, \bibinfo {author} {\bibfnamefont {E.~N.}\ \bibnamefont {Economou}}, \bibinfo {author} {\bibfnamefont {T.}~\bibnamefont {Koschny}}, \ and\ \bibinfo {author} {\bibfnamefont {C.~M.}\ \bibnamefont {Soukoulis}},\ }\bibfield  {title} {\enquote {\bibinfo {title} {Unifying approach to left--handed material design},}\ }\href@noop {} {\bibfield  {journal} {\bibinfo  {journal} {Opt.\ Lett.}\ }\textbf {\bibinfo {volume} {31}},\ \bibinfo {pages} {3620--3622} (\bibinfo {year} {2006})}\BibitemShut {NoStop}%
\bibitem [{\citenamefont {Wakatsuchi}\ \emph {et~al.}(2012)\citenamefont {Wakatsuchi}, \citenamefont {Paul}, \citenamefont {Greedy},\ and\ \citenamefont {Christopoulos}}]{MyCWeqCircuitPaper}%
  \BibitemOpen
  \bibfield  {author} {\bibinfo {author} {\bibfnamefont {H.}~\bibnamefont {Wakatsuchi}}, \bibinfo {author} {\bibfnamefont {J.}~\bibnamefont {Paul}}, \bibinfo {author} {\bibfnamefont {S.}~\bibnamefont {Greedy}}, \ and\ \bibinfo {author} {\bibfnamefont {C.}~\bibnamefont {Christopoulos}},\ }\bibfield  {title} {\enquote {\bibinfo {title} {Cut--wire metamaterial design based on simplified equivalent circuit models},}\ }\href@noop {} {\bibfield  {journal} {\bibinfo  {journal} {IEEE Trans.\ Antennas Propag.}\ }\textbf {\bibinfo {volume} {60}},\ \bibinfo {pages} {3670--3678} (\bibinfo {year} {2012})}\BibitemShut {NoStop}%
\bibitem [{\citenamefont {Imai}\ \emph {et~al.}(2023)\citenamefont {Imai}, \citenamefont {Homma}, \citenamefont {Takimoto}, \citenamefont {Tanikawa}, \citenamefont {Nakamura}, \citenamefont {Kaneko}, \citenamefont {Osaki}, \citenamefont {Niitsu}, \citenamefont {Cheng}, \citenamefont {Fathnan} \emph {et~al.}}]{imai2023design}%
  \BibitemOpen
  \bibfield  {author} {\bibinfo {author} {\bibfnamefont {S.}~\bibnamefont {Imai}}, \bibinfo {author} {\bibfnamefont {H.}~\bibnamefont {Homma}}, \bibinfo {author} {\bibfnamefont {K.}~\bibnamefont {Takimoto}}, \bibinfo {author} {\bibfnamefont {M.}~\bibnamefont {Tanikawa}}, \bibinfo {author} {\bibfnamefont {J.}~\bibnamefont {Nakamura}}, \bibinfo {author} {\bibfnamefont {M.}~\bibnamefont {Kaneko}}, \bibinfo {author} {\bibfnamefont {Y.}~\bibnamefont {Osaki}}, \bibinfo {author} {\bibfnamefont {K.}~\bibnamefont {Niitsu}}, \bibinfo {author} {\bibfnamefont {Y.}~\bibnamefont {Cheng}}, \bibinfo {author} {\bibfnamefont {A.~A.}\ \bibnamefont {Fathnan}},  \emph {et~al.},\ }\bibfield  {title} {\enquote {\bibinfo {title} {Design and analysis for the spice parameters of waveform-selective metasurfaces varying with the incident pulse width at a constant oscillation frequency},}\ }\href@noop {} {\bibfield  {journal} {\bibinfo  {journal} {Sci.\ Rep.}\ }\textbf {\bibinfo {volume} {13}},\ \bibinfo {pages} {7202} (\bibinfo
  {year} {2023})}\BibitemShut {NoStop}%
\bibitem [{\citenamefont {Asano}, \citenamefont {Nakasha},\ and\ \citenamefont {Wakatsuchi}(2020)}]{aplEqCircuit4WSM}%
  \BibitemOpen
  \bibfield  {author} {\bibinfo {author} {\bibfnamefont {K.}~\bibnamefont {Asano}}, \bibinfo {author} {\bibfnamefont {T.}~\bibnamefont {Nakasha}}, \ and\ \bibinfo {author} {\bibfnamefont {H.}~\bibnamefont {Wakatsuchi}},\ }\bibfield  {title} {\enquote {\bibinfo {title} {Simplified equivalent circuit approach for designing time-domain responses of waveform-selective metasurfaces},}\ }\href@noop {} {\bibfield  {journal} {\bibinfo  {journal} {Appl. Phys. Lett.}\ }\textbf {\bibinfo {volume} {116}},\ \bibinfo {pages} {171603} (\bibinfo {year} {2020})}\BibitemShut {NoStop}%
\bibitem [{\citenamefont {Fathnan}\ \emph {et~al.}(2022)\citenamefont {Fathnan}, \citenamefont {Homma}, \citenamefont {Sugiura},\ and\ \citenamefont {Wakatsuchi}}]{fathnan2022method}%
  \BibitemOpen
  \bibfield  {author} {\bibinfo {author} {\bibfnamefont {A.~A.}\ \bibnamefont {Fathnan}}, \bibinfo {author} {\bibfnamefont {H.}~\bibnamefont {Homma}}, \bibinfo {author} {\bibfnamefont {S.}~\bibnamefont {Sugiura}}, \ and\ \bibinfo {author} {\bibfnamefont {H.}~\bibnamefont {Wakatsuchi}},\ }\bibfield  {title} {\enquote {\bibinfo {title} {Method for extracting the equivalent admittance from time-varying metasurfaces and its application to self-tuned spatiotemporal wave manipulation},}\ }\href@noop {} {\bibfield  {journal} {\bibinfo  {journal} {J.\  Phys.\ D: Appl.\ Phys.}\ }\textbf {\bibinfo {volume} {55}},\ \bibinfo {pages} {015304} (\bibinfo {year} {2022})}\BibitemShut {NoStop}%
\bibitem [{\citenamefont {Ozawa}\ and\ \citenamefont {Wakatsuchi}(2025)}]{ozawa2025experimental}%
  \BibitemOpen
  \bibfield  {author} {\bibinfo {author} {\bibfnamefont {K.}~\bibnamefont {Ozawa}}\ and\ \bibinfo {author} {\bibfnamefont {H.}~\bibnamefont {Wakatsuchi}},\ }\bibfield  {title} {\enquote {\bibinfo {title} {Experimental validation of the diverse incident angle performance of a pulse-width-dependent antenna based on a waveform-selective metasurface in a reverberation chamber},}\ }\href@noop {} {\bibfield  {journal} {\bibinfo  {journal} {AIP Adv.}\ }\textbf {\bibinfo {volume} {15}} (\bibinfo {year} {2025})}\BibitemShut {NoStop}%
\bibitem [{\citenamefont {Munk}(2000)}]{MunkBook}%
  \BibitemOpen
  \bibfield  {author} {\bibinfo {author} {\bibfnamefont {B.~A.}\ \bibnamefont {Munk}},\ }\href@noop {} {\emph {\bibinfo {title} {Frequency selective surfaces: theory and design}}}\ (\bibinfo  {publisher} {A Wiley--Interscience Publication},\ \bibinfo {address} {New York, NY},\ \bibinfo {year} {2000})\BibitemShut {NoStop}%
\bibitem [{\citenamefont {Wakatsuchi}\ and\ \citenamefont {Christopoulos}(2011)}]{cwFiltering}%
  \BibitemOpen
  \bibfield  {author} {\bibinfo {author} {\bibfnamefont {H.}~\bibnamefont {Wakatsuchi}}\ and\ \bibinfo {author} {\bibfnamefont {C.}~\bibnamefont {Christopoulos}},\ }\bibfield  {title} {\enquote {\bibinfo {title} {Generalized scattering control using cut--wire--based metamaterials},}\ }\href@noop {} {\bibfield  {journal} {\bibinfo  {journal} {Appl.\ Phys.\ Lett.}\ }\textbf {\bibinfo {volume} {98}},\ \bibinfo {pages} {{221105}} (\bibinfo {year} {2011})}\BibitemShut {NoStop}%
\bibitem [{\citenamefont {Yang}\ \emph {et~al.}(2012)\citenamefont {Yang}, \citenamefont {Liu}, \citenamefont {Zhou},\ and\ \citenamefont {Cui}}]{yang2012reduction}%
  \BibitemOpen
  \bibfield  {author} {\bibinfo {author} {\bibfnamefont {X.~M.}\ \bibnamefont {Yang}}, \bibinfo {author} {\bibfnamefont {X.~G.}\ \bibnamefont {Liu}}, \bibinfo {author} {\bibfnamefont {X.~Y.}\ \bibnamefont {Zhou}}, \ and\ \bibinfo {author} {\bibfnamefont {T.~J.}\ \bibnamefont {Cui}},\ }\bibfield  {title} {\enquote {\bibinfo {title} {Reduction of mutual coupling between closely packed patch antennas using waveguided metamaterials},}\ }\href@noop {} {\bibfield  {journal} {\bibinfo  {journal} {IEEE Antennas Wirel.\ Propag.\ Lett.}\ }\textbf {\bibinfo {volume} {11}},\ \bibinfo {pages} {389--391} (\bibinfo {year} {2012})}\BibitemShut {NoStop}%
\bibitem [{\citenamefont {Wakatsuchi}(2015)}]{wakatsuchi2015time}%
  \BibitemOpen
  \bibfield  {author} {\bibinfo {author} {\bibfnamefont {H.}~\bibnamefont {Wakatsuchi}},\ }\bibfield  {title} {\enquote {\bibinfo {title} {Time-domain filtering of metasurfaces},}\ }\href@noop {} {\bibfield  {journal} {\bibinfo  {journal} {Sci. Rep.}\ }\textbf {\bibinfo {volume} {5}},\ \bibinfo {pages} {16737} (\bibinfo {year} {2015})}\BibitemShut {NoStop}%
\end{thebibliography}

\providecommand{\noopsort}[1]{}\providecommand{\singleletter}[1]{#1}%

\end{document}